\documentclass{jfm}

\usepackage{graphicx}
\usepackage{natbib}
\usepackage{amsmath,epsfig}
\usepackage{color}

\ifCUPmtlplainloaded \else
  \checkfont{eurm10}
  \iffontfound
    \IfFileExists{upmath.sty}
      {\typeout{^^JFound AMS Euler Roman fonts on the system,
                   using the 'upmath' package.^^J}%
       \usepackage{upmath}}
      {\typeout{^^JFound AMS Euler Roman fonts on the system, but you
                   dont seem to have the}%
       \typeout{'upmath' package installed. JFM.cls can take advantage
                 of these fonts,^^Jif you use 'upmath' package.^^J}%
      }
  \else
  \fi
\fi

\ifCUPmtlplainloaded \else
  \checkfont{msam10}
  \iffontfound
    \IfFileExists{amssymb.sty}
      {\typeout{^^JFound AMS Symbol fonts on the system, using the
                'amssymb' package.^^J}%
       \usepackage{amssymb}%
       \let\le=\leqslant  
       \let\ge=\geqslant  
      }{}
  \fi
\fi

\ifCUPmtlplainloaded \else
  \IfFileExists{amsbsy.sty}
    {\typeout{^^JFound the 'amsbsy' package on the system, using it.^^J}%
     \usepackage{amsbsy}}
    {\providecommand\boldsymbol[1]{\mbox{\boldmath $##1$}}}
\fi

\newcommand\crule[3][black]{\textcolor{#1}{\rule{#2}{#3}}}

\providecommand\bnabla{\boldsymbol{\nabla}}

\newsavebox{\astrutbox}
\sbox{\astrutbox}{\rule[-5pt]{0pt}{20pt}}

\def\build#1_#2^#3{\mathrel{\mathop{\kern 0pt#1}\limits_{#2}^{#3}}}
\def \E{ \mathbb E  }
\def \R{ \mathbb R  }

\newcommand\refp[1]{(\ref{#1})}
\newcommand\tr{\mbox{tr}}
\newcommand\bs[1]{\boldsymbol{#1}}
\newcommand\bu{\boldsymbol{u}}
\newcommand\bx{\boldsymbol{x}}
\newcommand\by{\boldsymbol{y}}

\newcommand\bh{\boldsymbol{h}}
\newcommand\bp{\boldsymbol{p}}
\newcommand\bo{\boldsymbol{\omega}}

\newcommand\tA{\mathsfbi{A}}
\newcommand\tS{\mathsfbi{S}}
\newcommand\tO{\mathsfbi{\Omega}}
\newcommand\tV{\mathsfbi{V}}
\newcommand\tI{\mathsfbi{I}}
\newcommand\tC{\mathsfbi{C}}

\newcommand\tW{\mathsfbi{W}}
\newcommand\tNo{\mathsfbi{\mathcal N}}
\newcommand\modcirc{\protect\raisebox{-1.1pt}{\protect\scalebox{1.7}{$\circ$}}}
\newcommand\modsquare{\crule{0.2cm}{0.2cm}}

\title[A multifractal model for the velocity gradients dynamics in turbulent flows]{A multifractal model for the velocity gradients dynamics in turbulent flows}

\author[ ]%
{Rodrigo M. Pereira$^{1}$\thanks{rpereira@df.ufpe.br}, Luca Moriconi$^{2}$, Laurent Chevillard$^3$}

\affiliation{$^1$Laborat\'orio de F\'isica Te\'orica e Computacional, Departamento de F\'isica, Universidade Federal de Pernambuco, 50670-901, Recife, PE, Brazil\\
$^2$Instituto  de  F\'isica, Universidade Federal do Rio de Janeiro, C.P. 68528, 21945-970, Rio  de  Janeiro,  RJ,  Brazil\\
$^3$Univ Lyon, Ens de Lyon, Univ Claude Bernard, CNRS, Laboratoire de Physique, 46 all\'ee d'Italie F-69342 Lyon, France}
\pubyear{2018}
\volume{..}
\pagerange{..}
\date{..}

\begin{document}

\maketitle
\begin{abstract}

We develop a stochastic model for the velocity gradients dynamics along a Lagrangian trajectory in isotropic and homogeneous turbulent flows. Comparing with different attempts proposed in the literature, the present model, at the cost of introducing a free parameter known in turbulence phenomenology as the intermittency coefficient, gives a realistic picture of velocity gradient statistics at any Reynolds number. To achieve this level of accuracy, we use as a first modelling step a regularized self-stretching term in the framework of the Recent Fluid Deformation (RFD) approximation that was shown to give a realistic picture of small scales statistics of turbulence only up to moderate Reynolds numbers. As a second step, we constrain the dynamics, in the spirit of \cite{GirPop90}, in order to impose a peculiar statistical structure to the dissipation seen by the Lagrangian particle. This probabilistic closure uses as a building block a random field that fulfils the statistical description of the intermittency, i.e. multifractal, phenomenon. To do so, we define and generalize to a statistically stationary framework a proposition made by \cite{Sch03}. These considerations lead us to propose a non-linear and non-Markovian closed dynamics for the elements of the velocity gradient tensor. We numerically integrate this dynamics and observe that a stationary regime is indeed reached, in which (i) the gradients variance is proportional to the Reynolds number, (ii) gradients are typically correlated over the (small) Kolmogorov time scale and gradients norms over the (large) integral time scale (iii) the joint probability distribution function of the two non vanishing invariants $Q$ and $R$ reproduces the characteristic teardrop shape, (iv) vorticity gets preferentially aligned with the intermediate eigendirection of the deformation tensor and (v) gradients are strongly non-Gaussian and intermittent, a behaviour that we quantify by appropriate high order moments. Additionally, we examine the problem of rotation rate statistics of (axisymmetric) anisotropic particles as observed in Direct Numerical Simulations. Although our realistic picture of velocity gradient fluctuations leads to better results when compared to the former RFD approximation, it is still unable to provide an accurate description for the rotation rate variance of oblate spheroids.

\end{abstract}

\normalsize

%%%%%%%%%%%%%%%%%%%%%%%%%%%%%%%%%%%%%%%%%%%%%%%%%%%%%%%%%%%%%%%%%%%%%%%%%%%%%%
%%%%%%%%%%%%%%%%%%%%% Here the document begins %%%%%%%%%%%%%%%%%%%%%%%%%%%%%%%

%%%%%%%%%%%%%%%%%%%%%%%%%%%%%%%%%%%%%%%%%%%%%%%%%%%%%%%%%%%%%%%%%%%%%%%%%%%%%%
\section{Introduction}
%%%%%%%%%%%%%%%%%%%%%%%%%%%%%%%%%%%%%%%%%%%%%%%%%%%%%%%%%%%%%%%%%%%%%%%%%%%%%%%%%%%%%%%%%%%%%%%%%%%%%%%%%%%%%%%%%%%%%%%%%%%%%%%%%%%%%%%%%%%%%%%%%%%%%%%%%%%%%%%%%%%%%%%%%%%%%%%%%%%%%%%%%%%%%%%%%%%%%%%%%%%%%%

The study of the statistical properties of the velocity gradients dynamics has shed a new light on hydrodynamics turbulence. In the past years \citep{Tsi01,Wal09,Men11}, quantitative progress has been made while focusing on the small scales of turbulence, i.e. below the Kolmogorov length scale, in particular on the statistical and geometrical properties of the velocity gradient tensor $\tA$, defined as the $3\times 3$ matrix made up of the gradients of the three components of velocity $u_i$, for $1\le i \le 3$ along the three spatial directions $x_j$, with $1\le j\le 3$, namely
\begin{equation}\label{eq:DefA}
A_{ij} = \frac{\partial u_i}{\partial x_j}.
\end{equation}
Henceforth, as it is classically used in fluid mechanics, we will denote by $\tS$ and $\tO$ the symmetric and antisymmetric decompostion of $\tA$, respectively, such that $\tA=\tS + \tO$. It has been recognized that many of the geometrical properties of $\tA$, such as the relative amplitudes and signs of the eigenvalues of $\tS$ and the alignments of the vorticity vector $\bo = \bnabla \wedge \bu$ (related to $\tO$ according to $\tO\bh = \frac{1}{2}\bo\wedge \bh$ for all vectors $\bh \in \R^3$) with respect to the eigenframe of $\tS$, can be understood in a kinematic sense while considering the Lagrangian evolution of $\tA$. This dynamics is obtained by taking a spatial derivative of the Navier-Stokes equations and reads
\begin{equation}\label{eq:NSA}
\frac{dA_{ij}}{dt} = -A_{ik}A_{kj} - \frac{\partial^2 p}{\partial x_i\partial x_j}+\nu \Delta A_{ij},
\end{equation}
where $d/dt \equiv \partial/\partial t + \bu\cdot\bnabla$ stands for the material derivative, $p$ is the pressure field determined by the incompressibility condition, hence solution of the respective Poisson equation $\Delta p = -\tr(\tA^2)$, and $\nu$ the kinematic viscosity \citep{Tsi01,Wal09,Men11}.

The transport equation \refp{eq:NSA} of the velocity gradient tensor $\tA$ states that its time variation along a Lagrangian trajectory is governed by the action of three terms: the self-stretching term $-\tA^2$, the pressure Hessian and the viscous diffusion. Let us mention that \refp{eq:NSA} is not closed in terms of the temporal profile of the nine elements of $\tA$ along the trajectory of the fluid particle under consideration since both the pressure and viscous terms require the knowledge of the whole spatial field. Indeed, the viscous term comprises the gradients of velocity gradients, which are not known along a single trajectory, and it is known \citep{Ohk93,CheMen08,WilMen14} that the pressure Hessian can be expressed as a convolution over space of $\tr(\tA^2)$, as a consequence of the Poisson equation. It is nonetheless tempting to focus on a trajectory of a single particle, disregarding all the others, thus decreasing drastically the number of degrees of freedom. This implies studying an approximative dynamics based on simplistic closures of the pressure Hessian and viscous terms that are, hopefully, realistic enough to reproduce the observed statistics of $\tA$ in experimental and numerical flows.

In this spirit, by neglecting the anisotropic part of the pressure Hessian entering \refp{eq:NSA} and taking $\nu=0$ one builds the simplest closure that preserves incompressibility. This is the so-called restricted Euler (RE) approximation \citep[][see also \citealt{Men11} for a review]{Vie82,Can92}. It is known that this closure leads to a finite time singularity of the elements of $\tA$ for any non-vanishing initial condition. This predicted singularity is unphysical, although this closure allows to understand, among others, one experimental observation: vorticity has the tendency to align with the eigendirection of the deformation $\tS$ associated to the intermediate eigenvalue. The RE approximation is thus an appealing starting point to design more realistic closures of pressure and diffusivity able, at least, to regularize the finite time singularity implied by the self-stretching term, while keeping track of the underlying non-linear dynamics. Closures can then be compared against experimental measurements and numerical simulations. Usually, this closed dynamics is associated with a tensorial random forcing, delta-correlated in time, to maintain a statistically steady state \citep{Men11}. In the stationary regime, the statistics of $\tA$ can then be compared to those obtained, for instance, in numerical simulations of homogeneous and isotropic turbulence, available at several databases in the world  \citep[see for instance][]{LiPer08}. Recently, several such closures for the pressure Hessian and the viscous term have been proposed in the literature \citep{CheMen06,WilMen14,JohMen16}. They all show realistic statistics of, among other properties, the joint probability distribution function (PDF) of the two invariants $Q$ and $R$, and the peculiar preferential alignment of vorticity with the eigenframe of deformation (we will define precisely these quantities later in the article). Moreover, with different levels of success, they even compared well against more precise estimations of the pressure field, such as conditional averages of the pressure Hessian given an instance of $\tA$, and more precisely, of a joint instance of $R$ and $Q$ \citep{CheMen08,Men11,CheLev11,WilMen14,JohMen16}. Whereas all these closed dynamics have in common to preserve the self stretching term $-\tA^2$, they unfortunately, to our knowledge, give unrealistic statistics when the Reynolds number $\mathcal R_e$, introduced as a natural free parameter of the closures, becomes too large, preventing the model from reaching the asymptotic limit of infinite Reynolds numbers. They nonetheless provide realistic closures at moderate Reynolds numbers, and they are at least able to regularize the finite-time singularity implied by the self-stretching term $-\tA^2$. To this regards, they appear as good starting points to build up more sophisticated models that can describe flows at arbitrarily large Reynolds Numbers. Let us now define more precisely the underlying dynamics that we will study. Consider the following stochastic differential equation for the velocity gradient tensor $\tA$
\begin{equation}\label{eq:SDE}
dA_{ij} = V_{ij}dt +D_{ijkl} W_{kl}(dt),
\end{equation}
where $V_{ij}$ and $D_{ijkl}$ are called respectively the \emph{drift} and \emph{diffusion} terms, and $W_{kl}(dt)$ is a realization of a white noise stochastic process, defined such that all components are independent and, loosely speaking, delta-correlated in time. Let us assume at this stage that, following this dynamics \refp{eq:SDE}, $\tA$ reaches a statistically stationary regime, characterized for example by the covariance of its elements, supposed thus finite. We are then asking the tensor terms $V_{ij}$ and $D_{ijkl}$ from \refp{eq:SDE} to be, as a necessary condition, also of finite variance.
In the framework of additive noise, the diffusion term is deterministic, and the covariance of its elements can be chosen such that it is consistent with a fourth-order isotropic tensor, ensuring furthermore that the trace-free condition imposed by incompressibility is respected \citep{CheMen08,JohMen16}. As for the drift term $V_{ij}$, let us for example consider the one predicted by the restricted Euler (RE) approximation, that reads in matrix form
\begin{equation}\label{eq:REDeterm}
\tV^{\mbox{\tiny{RE}}} = -\tA^2+\frac{\tr(\tA^2)}{3}\tI,
\end{equation}
where $\tI$ stands for the identity matrix, and the one constructed under the Recent Fluid Deformation (RFD) approximation, namely
\begin{equation}\label{eq:RFDDeterm}
\tV^{\mbox{\tiny{RFD}}} = -\tA^2+\frac{\tr(\tA^2)}{\tr(\tC_{\tau_\eta}^{-1})}\tC_{\tau_\eta}^{-1} -\frac{\tr(\tC_{\tau_\eta}^{-1})}{3T}\tA,
\end{equation}
where $\tau_\eta$ and $T$ are respectively the Kolmogorov and integral time scales, the two characteristic scales of turbulence at a given Reynolds number $\mathcal R_e$, and $\tC_{\tau_\eta}$ the short-time Cauchy-Green tensor
\begin{equation}\label{eq:CGtaueta}
\tC_{\tau_\eta} = e^{\tau_\eta\tA}e^{\tau_\eta\tA^\top}.
\end{equation}
The RFD approximation and its consequences have been discussed in \cite{CheMen06,CheMen08,Men11}, and we will not recall them here. We will keep in mind that whereas the RE approximation leads to finite time singularity and is independent of the Reynolds number, the RFD approximation regularizes this finite-time singularity through the joint action of the modelled pressure Hessian and viscous term using the short time Cauchy-Green tensor $\tC_{\tau_\eta}$ \refp{eq:CGtaueta}. Furthermore, the Reynolds number becomes explicit through the ratio of the two time scales $T/\tau_\eta \propto \sqrt{\mathcal R_e}$. Similar drift terms can be derived from the approaches developed in \cite{WilMen14} and \cite{JohMen16}. As commented before, such closures give a realistic picture of the statistics of $\tA$ at moderate Reynolds numbers, i.e. when the ratio $\tau_\eta/T$ does not become too small. As an example, the RFD approximation gives realistic results of homogeneous and isotropic turbulence (of Taylor-based Reynolds number $\mathcal R_\lambda=140$) when $\tau_\eta/T=0.1$, and deteriorate for $\tau_\eta/T<0.01$ \citep{CheMen08}. Subsequent numerical and theoretical developments in the framework of the RFD approximation indeed show that the model pointed in \refp{eq:RFDDeterm} fails at giving a realistic picture of turbulence at high Reynolds numbers \citep{AfoMen10,MorPer14,GriBou17}. We may wonder whether it is possible to constrain the drift $V_{ij}$ and diffusion $D_{ijkl}$ terms in a different way to ensure, for instance, a basic property of turbulence \citep{Fri95} that is the finiteness of the average dissipation rate $2\nu \E \left[\tr(\tS^2)\right]$ when $\mathcal R_e\to \infty$.

Actually, such an approach has already been explored in this context by \cite{GirPop90}. They end up with a very different dynamics for $\tA$ (in the spirit of \refp{eq:SDE}), where a diffusion term $D_{ijkl}$ that depends explicitly on $\tA$ appears, thus lying in the class of multiplicative noise models. More precisely, $V_{ij}$ and $D_{ijkl}$ are constrained to ensure that a particular contraction of $\tA$, namely the pseudo-dissipation $\varphi = \tr(\tA\tA^\top)$ (in units of viscosity $\nu$), follows the dynamics of an exponentiated Ornstein-Uhlenbeck (OU) process. They thus impose that the dynamics of $\tA$ \refp{eq:SDE}, in particular, fulfils the dimensional Kolmogorov's prediction
\begin{equation}\label{eq:KolPredPD}
\E(\varphi) = \E\left[\tr(\tA\tA^\top)\right] = \frac{1}{\tau_\eta^2},
\end{equation}
which thus defines precisely what we mean by the Kolmogorov time scale $\tau_\eta$. Consequently, imposing the condition \refp{eq:KolPredPD}, they are able to reach any Reynolds numbers, as far as the variance of the elements of $\tA$ are concerned. But assuming that $\varphi$ is an exponentiated OU process has further consequences. In particular, $\log(\varphi)$ is thus a Gaussian process, whose average is set by fulfilling the Kolmogorov prediction given in \refp{eq:KolPredPD}, and whose covariance decays exponentially along Lagrangian trajectories over a characteristic time scale. This was already checked in early direct numerical simulations of Lagrangian turbulence \citep{YeuPop89}, where it is observed that indeed, to a good approximation, $\log (\varphi)$ is a Gaussian random process whose covariance is consistent with an exponential decrease over the large time scale $T$ (or integral turnover time scale) of turbulence. This stochastic modelling  of $\varphi$ compares thus well at this stage to numerical flows and is defined up to a free parameter, called $\hat{a}$ in \cite{GirPop90}, that has to be determined from empirical data. Unfortunately, higher order moments of $\varphi$ are poorly predicted, and do not compare appropriately with the $\mathcal R_e$ dependence of the fluctuations observed in experimental \citep{VanAnt80,SreAnt97} and numerical \citep{IshKan07} flows. We will develop these ideas in section \ref{Sec:GPDiss}. Let us just keep in mind that this modelling  needs to be improved in order to be realistic in face of observed fluctuations. To this regard, it turns out that the $\mathcal R_e$-dependence of moments of velocity derivatives observed experimentally and numerically \citep{SreAnt97,IshKan07} are well described by the multifractal formalism, as depicted, for instance, in \cite{Bor93} and \cite{Fri95}. We need thus to develop a new stochastic model for pseudo-dissipation able to reproduce the observed fluctuations of the velocity gradients, and furthermore to be consistent with the multifractal picture.

Such a random field, able to reproduce the statistics of the pseudo-dissipation, is at the core of the work of \cite{Kol62} and \cite{Obu62}, where the dissipation field, and alternatively, the pseudo-dissipation field, determine the statistical features of the velocity increments through the refined similarity hypothesis \citep{Fri95}. In an Eulerian description of the flow, the \textit{spatial} field $\varphi$ is modelled in a homogeneous way by taking the exponential of a Gaussian field, whose covariance decreases logarithmically over the integral length scale $L$. This stochastic construction was first proposed by \cite{Man72} following a construction due to \cite{Yag66}, and formalised by \cite{Kah85} in the context of Multiplicative Chaos Theory \citep{RhoVar14}. Going back to the velocity gradient dynamics where the Lagrangian point of view has to be adopted, we need a temporal form of such a multiplicative chaos. This was considered by \cite{Sch03}, where, as a first step, it is shown how to adapt the approach of \cite{Kah85} in a Lagrangian context that requires, in particular, a causal construction. \cite{Sch03} advances also some heuristics in order to build a dynamics, i.e. a stochastic differential equation, for which the statistically stationary solution is indeed such a causal multifractal process. We will develop these ideas in section \ref{Sec:MultiDiss}.

The purpose of the present article is to include the multifractal picture given by the formerly described multiplicative chaos approach into the dynamics of the velocity gradient tensor \refp{eq:SDE}. In the sequel, we will develop this idea and finally propose the following stochastic model for $\tA$ along a Lagrangian trajectory
\begin{equation}\label{eq:DynAFinalIntro}
dA_{ij} = \left[V^{\mbox{\tiny{RFD}}}_{ij}+f(t)A_{ij}\right]dt + \frac{1}{2}\sqrt{\frac{\mu^l\varphi}{\tau_\eta}}\left[W_{ij}-\frac{1}{3}\tr(\tW)\delta_{ij}\right].
\end{equation}
Let us now comment on the various terms entering the proposed dynamics \refp{eq:DynAFinalIntro}. First, the \textit{drift} term proportional to $dt$ is made up of the dynamics given by the RFD closure \refp{eq:CGtaueta}, supplemented by an additional damping term proportional to $\tA$ itself. The scalar function $f(t)$, given in \refp{eq:ExpressfA}, depends explicitly on $\tA$ at present time $t$ and on an additional random variable (given in \refp{eq:betahat}) built on the past values of both $\tA$ and the white noise tensor field $\tW$. Its presence is crucial to ensure multifractal properties in the stationary regime, and makes the overall dynamics \refp{eq:DynAFinalIntro} non-Markovian. Secondly, the \textit{diffusion} term is very similar to the one used by \cite{GirPop90} that, as mentioned, have a multiplicative nature. It is given by the product of the trace-free part of the white tensor noise $\tW(dt)$ and the pseudo-dissipation $\varphi = \tr(\tA\tA^\top)$ evaluated at present time $t$, and its intensity is moderated by a free parameter $\mu^l$, interpreted as the intermittency coefficient from turbulence phenomenology, for reasons that will become clearer.

In section \ref{Sec:StoModDiss}, we review and recall the two existing stochastic models for the pseudo-dissipation seen by a fluid particle. This includes the proposition made by \cite{Pop90}, that is to take the exponential of an Ornstein-Uhlenbeck process, and the one of \cite{Sch03} that we define rigorously and extend in order to deal with statistically stationary processes. In section \ref{Sec:InclMCRFD}, following the idea of \cite{GirPop90}, we show how to include the multifractal picture into the dynamics of the velocity gradient tensor. We end up with a proposition for a closed tensorial stochastic differential equation \refp{eq:DynAFinalIntro}, which is non-linear and non-Markovian. Section \ref{sec:Num} is devoted to the description of the numerical procedure used to simulate the trajectories of our process \refp{eq:DynAFinalIntro}. We present then in section \ref{sec:NumRes} the results of numerical simulations of the trajectories for various Reynolds numbers and discuss their comparison with known empirical facts of turbulence. We gather in section \ref{Sec:Conclu} our conclusions and perspectives.

%%%%%%%%%%%%%%%%%%
%%  NEW SECTION %%
%%%%%%%%%%%%%%%%%%

\section{Stochastic models of pseudo-dissipation}\label{Sec:StoModDiss}

\subsection{Pseudo-dissipation as an exponentiated Ornstein-Uhlenbeck process}\label{Sec:GPDiss}

As a building block of their stochastic model for the velocity gradients, \cite{GirPop90} have considered for pseudo-dissipation $\varphi$ the exponential of an Ornstein-Uhlenbeck process. This model is justified by the observations made in early numerical simulations of \cite{YeuPop89} that the shape of the one point PDF of $\log\varphi$ is close to a Gaussian and, furthermore, the autocorrelation of $\log\varphi(t)$ follows an exponential decrease over a typical time scale given by the integral time scale $T$ of turbulence \citep{Pop90}. At this point, we may remark that the small scale quantity $\varphi$ is correlated over the large energy containing scale of turbulence, which is characteristic of the absence of scale decoupling of turbulence. This was already recognized in the Eulerian framework where it is observed that dissipation is correlated in space over the large integral length scale $L$ \citep{GagHop79,AntPha81}. From the theoretical side, this long-range correlated nature of the small scales is at the heart of the models of intermittency of \cite{Kol62} and \cite{Obu62} (see also \citealt{Nov89,Nov90} for discussions in the lagrangian framework), although, at that stage, nothing is said about how the correlation function decreases over $T$, and as we will see, multifractal phenomenology requires a logarithmic decrease (see subsection \ref{Sec:MultiDiss}) which is not reproduced by the proposition of \cite{Pop90}.

The statistical properties of this model are reviewed in appendix \ref{ann:ExpOU}. Its causal dynamics is given in \refp{eq:SDEGPDiss} and its unique solution, given an initial condition, is given explicitly in \refp{eq:ExpOUDiss}. Let us call $\hat{a}$, as in \cite{GirPop90}, the free parameter of this model. It can be conveniently related to the variance of the logarithm of pseudo-dissipation as $\E\log^2\varphi - \E^2\log\varphi = 2\hat{a}^2T$, or equivalently, to the moments $\E(\varphi^q)=\frac{1}{\tau_{\eta}^{2q}}e^{\hat{a}^2Tq(q-1)}$ \refp{eq:MomqExpOUDiss}. Whatever the value of $\hat{a}$, this model fulfils the basics of Kolmogorov's phenomenology, namely $\E(\varphi)=1/\tau_{\eta}^{2}$ \refp{eq:KolPredPD}. As it is explained and observed in \cite{YeuPop89}, $\hat{a}^2$ is expected to depend logarithmically on the Reynolds number (see \refp{eq:Choiceabar}) in order to be consistent with both empirical observations and the phenomenology of \cite{Kol62} and \cite{Obu62} such as $\hat{a}^2\sim \log \mathcal R_e$ when $\mathcal R_e\to \infty$.

To build a more precise connection with \cite{Kol62} and \cite{Obu62} phenomenology, and more generally with the multifractal (i.e. intermittent) phenomenology \citep{Bor93,Fri95}, let us consider the average value of pseudo-dissipation over a time interval of extension $\tau$, i.e.
\begin{equation}\label{eq:MeanTauDiss}
\varphi_\tau(t) = \frac{1}{\tau}\int_{s\in [t-\tau,t]}\varphi(s)ds.
\end{equation}
This quantity is the Lagrangian analogue of the (pseudo-)dissipation averaged over a ball of size $\ell$ as considered by \cite{Kol62} and \cite{Obu62} in the refined similarity hypothesis relating velocity structure functions and dissipation field \citep{Bor93,Fri95}.
It is shown in appendix \ref{ann:ExpOU} (see equation \ref{eq:VarPhiTauqOU}) how to compute the high-order moments of $\varphi_\tau$, and it is easy to be convinced that if $\hat{a}^2T$ is taken proportional to $\log \mathcal R_e$, as it is required by the intermittent phenomenology formerly described, then, for any $\tau>0$, we get diverging high-order moments of the coarse-grained pseudo-dissipation, i.e., for $q>1$, $\tau_\eta^{2q}\E[\varphi_\tau^q]\to \infty$ when $\mathcal R_e\to \infty$, which is not consistent with the refined similarity hypothesis (see the discussion on the Reynolds number dependence of the model in appendix \ref{ann:ExpOU}). Thus, taking as a stochastic model for pseudo-dissipation the exponential of an Ornstein-Uhlenbeck process prevents us from fulfilling the statistical properties required by the phenomenology of \cite{Kol62} and \cite{Obu62}. We will see in the following section that, on the other hand, a pseudo-dissipation model given by the multiplicative chaos allows one to obtain (i) moments $\E[\varphi^q]$ that scale as a power of the Reynolds number, and (ii) bounded moments $\varphi_\tau$ for any finite $\tau>0$ when $\mathcal R_e\to \infty$.

\subsection{Pseudo-dissipation as a causal and stationary multiplicative chaos}\label{Sec:MultiDiss}

Multiplicative chaos has been introduced by \cite{Man72} in the context of turbulence, following the work of \cite{Yag66}, and formalized by \cite{Kah85} in order to give a stochastic meaning to the dissipation field as depicted in the phenomenology of \cite{Kol62} and \cite{Obu62} (see recent mathematical developments in \cite{RhoVar14}). In the Eulerian framework, for which we need to define spatial fields, dissipation is taken as the exponential of a isotropic Gaussian field logarithmically correlated over the integral length scale $L$. As briefly reviewed in \cite{PerGar16}, we can give a clear meaning to the dissipation field as depicted in KO62 while proposing a stochastic representation of it. At a given finite Reynolds number, implying a finite Kolmogorov length scale $\eta$, this representation reads
\begin{equation}\label{eq:MCPDEul}
\varphi(\bx) = \frac{1}{\tau_\eta^2}e^{\sqrt{\mu^e}X_{\eta}(\bx)-\frac{\mu^e}{2}\E\left[X_{\eta}^2\right]},
\end{equation}
where $X_{\eta}$ is a homogeneous and isotropic Gaussian field whose variance diverges logarithmically with $\eta$ and is logarithmically correlated over the integral length scale $L$. Being Gaussian, the field $X_{\eta}$ can be conveniently expressed as a linear operation (i.e. a convolution) on a white measure as
\begin{equation}\label{eq:XetaEul}
X_{\eta}(\bx)=\frac{1}{4\pi}\int_{|\bx-\by|\le L}\frac{1}{|\bx-\by|_{\eta}^{3/2}}W(dy),
\end{equation}
where a regularized norm $|\bx|_\eta$ is introduced, and whose exact expression has little importance at this stage, since the statistical properties of $X_{\eta}$ become independent on its precise form in the limit of infinite Reynolds number (i.e. $\eta\to 0$). We will take for instance, to illustrate our purpose, the homogeneous and isotropic small-scale cut-off $|\bx|^2 = |\bx|^2+\eta^2$ for any $\bx\in\mathbb R^3$. See \cite{RhoVar14,PerGar16} for a discussion on this subject.

It is then possible to show (see for example \citealt{PerGar16} for detailed computations and \citealt{RhoVar14} for a general approach) that
\begin{equation}\label{eq:DivVarXEta}
 \E\left[X_{\eta}^2\right]\build{\sim}_{\eta\to 0}^{}\log \left( \frac{L}{\eta}\right).
 \end{equation}
Furthermore, locally, we have the convergence, for $|\bh|> 0$,
\begin{equation}\label{eq:LogCorrXEta}
 \lim_{\eta\to 0}\E\left[X_{\eta}(\bx)X_{\eta}(\bx+\bh)\right]= \log^+ \left( \frac{L}{|\bh|}\right) + g(|\bh|),
 \end{equation}
where $g$ is a continuous and bounded function of its argument, and $\log^+(|\bx|)=\log(|\bx|)$ for $|\bx|\ge 1$ and vanishes elsewhere. Doing so, we can derive the statistical properties of the modelled  (Eulerian) pseudo-dissipation field \refp{eq:MCPDEul} that fulfils the axiomatics of KO62, namely
\begin{equation}\label{eq:MomEulPDMC}
\E \left[\varphi^q\right]\build{\sim}_{\eta\to 0}^{}\frac{e^{\frac{\mu^e}{2}q(q-1) g(0)}}{\tau_\eta^{2q}}\left(\frac{L}{\eta}\right)^{\frac{\mu^e}{2}q(q-1)}.
\end{equation}
Recalling that the Kolmogorov scale $\eta$ behaves with the Reynolds number $\mathcal R_e$ as $\eta\propto L\mathcal R_e^{-3/4}$, we see that modelling the pseudo-dissipation as a multiplicative chaos \refp{eq:MCPDEul} gives moments proportional to a power of the Reynolds number \refp{eq:MomEulPDMC} as required by KO62. Furthermore, the locally averaged pseudo-dissipation
\begin{equation}\label{eq:MeanTauEll}
\varphi_\ell(\bx) = \frac{1}{\frac{4}{3}\pi\ell^3}\int_{|\bx-\by|\le\ell}\varphi(\by)dy,
\end{equation}
is such that its moments remain bounded when $\eta\to 0$ and we have the following behaviour at small scales
\begin{equation}\label{eq:MomMeanTauEll}
\lim_{\eta\to 0}\E \left[\varphi_\ell^q\right] \build{\sim}_{\ell\to 0}^{}C_q\left( \frac{\ell}{L}\right)^{\frac{\mu^e}{2}q(1-q)},
\end{equation}
where $C_q$ is a scale and Reynolds number independent constant, that can be eventually calculated.

We can see from the proposition of \cite{Man72} that taking the pseudo-dissipation \refp{eq:MCPDEul} as the exponential of a log-correlated field \refp{eq:XetaEul} ensures that its moments behave as power laws of the Reynolds number \refp{eq:MomEulPDMC}, in particular its average is consistent with Kolmogorov's prediction \refp{eq:KolPredPD}. Moreover, moments of its locally averaged version over a ball of radius $\ell$ \refp{eq:MeanTauEll} become independent of the Reynolds number and behave as power laws at small scales \refp{eq:MomMeanTauEll}.

In the Lagrangian context of velocity gradient dynamics we are interested in (equation \refp{eq:NSA}), we would like to build up a unidimensional version of the multiplicative chaos used in the Eulerian framework \refp{eq:MCPDEul} to model pseudo-dissipation as seen by a fluid particle along its trajectory. It has to be stationary, its variance particularly should be time independent, and, furthermore, it has to be causal. Causality is crucial here, not only as a physical requirement but, technically speaking, because the so-obtained multiplicative chaos has to be defined as a solution of a stochastic differential equation that will be coupled to the dynamics of $\tA$ \refp{eq:SDE}, as done by \cite{GirPop90} while imposing that $\tr(\tA\tA^\top)$ follows the statistics of an exponentiated Ornstein-Uhlenbeck process, as already depicted in section \ref{Sec:GPDiss}. Thus, instead of the three-dimensional spatial field considered in \refp{eq:MCPDEul}, we will consider the following one-dimensional temporal stochastic field
\begin{equation}\label{eq:MCPDLag}
\varphi(t) = \frac{1}{\tau_\eta^2}e^{\sqrt{\mu^l}X_{\tau_\eta}(t)-\frac{\mu^l}{2}\E\left[X_{\tau_\eta}^2\right]},
\end{equation}
where now $X_{\tau_\eta}(t)$ is a zero-average causal and stationary Gaussian field whose variance $\E[X_{\tau_\eta}^2]$ blows up logarithmically with the Kolmogorov time scale $\tau_\eta$ (instead of \refp{eq:DivVarXEta}), and whose covariance is independent of the Reynolds number and decreases logarithmically over the integral time scale $T$ of turbulence (in contrast to \refp{eq:LogCorrXEta}). As proved in appendix \ref{Ann:MultiChaosDiss}, the unique solution $X_{\tau_\eta}(t)$ of the following stochastic dynamics fulfils these two requirements
\begin{equation}\label{eq:sdeXText}
dX_{\tau_\eta}(t) = \left[-\frac{1}{T} X_{\tau_\eta}(t)+\beta_{\tau_\eta}(t)\right]dt + \frac{1}{\sqrt{{\tau_\eta}}}W(dt),
\end{equation}
being $W$ a Gaussian white noise and $\beta_{\tau_\eta}(s)$ a random function known in the literature as a (${\tau_\eta}$-regularized) fractional Gaussian noise of Hurst exponent $H=0$ \citep{ManVan68}
\begin{equation}\label{eq:betaText}
\beta_{\tau_\eta}(t) = -\frac{1}{2} \int_{s=-\infty}^t \frac{1}{(t-s+{\tau_\eta})^{3/2}} W(ds).
\end{equation}
We note that \refp{eq:sdeXText} is linear and so can be integrated in a simple fashion such that the solution  $X_{\tau_\eta}(t)$ can be equivalently defined as
\begin{equation}\label{eq:XText}
X_{\tau_\eta}(t) = \int_{s=-\infty}^t e^{-\frac{t-s}{T}}\beta_{\tau_\eta}(s)ds + \frac{1}{\sqrt{{\tau_\eta}}}\int_{s=-\infty}^t e^{-\frac{t-s}{T}}W(ds),
\end{equation}
where we have chosen formally the initial condition $X_{\tau_\eta}(-\infty)=0$. As we can see, the process $X_{\tau_\eta}$ is the sum of two Gaussian causal fields. The first process on the RHS of \refp{eq:XText} is similar, in a certain sense, to the one in the formulation of the Eulerian modelling of pseudo-dissipation \refp{eq:XetaEul}. It can be seen as an exponentially smoothed version of a fractional Brownian noise of vanishing Hurst exponent, and is known in the literature as a fractional Ornstein Uhlenbeck process \citep{CheKaw03}. The second process on the RHS of \refp{eq:XText} is a usual Ornstein-Uhlenbeck process whose variance blows up as $1/\tau_\eta$. Notice that the same white measure $W$ appears in both these Gaussian fields, making them correlated. This correlation will be responsible for a strong cancellation taking place in between these two processes. Actually, this cancellation, formalized in appendix \ref{Ann:MultiChaosDiss}, will make the variance of $X_{\tau_\eta}\sim \log (T/\tau_\eta)$ way smaller than the one of each of these formerly described Gaussian processes. The idea of adding a correlated process to a fractional Gaussian noise can be understood in practice with the heuristics developed in \cite{Sch03}. Our contribution to this matter is to properly regularize the dynamics proposed by \cite{Sch03}, injecting the large scale $T$ dependence not in the fractional Brownian noise, but through a linear damping, allowing us to rigorously define $X_{\tau_\eta}$ (appendix \ref{Ann:MultiChaosDiss}).

It is then possible to show that the pseudo-dissipation defined as the exponential of $X_{\tau_\eta}$ fulfils the KO62 requirements, namely, high order moments behave as a power laws of the Reynolds number
\begin{align}\label{eq:PredMomqLag}
\E \left[\varphi^q\right] \build{\propto}_{\tau_\eta\to 0}^{} \left(\frac{T}{\tau_\eta}\right)^{\frac{\mu^l}{2}q(q-1)},
\end{align}
where the multiplicative constant (given explicitly in \refp{eq:AnnMomqPhitaueta}) is independent of the Reynolds number, and moments of its coarse-grained version along the trajectory \refp{eq:MeanTauDiss} behaving as power laws of the time scale $\tau$ in the asymptotic limit of infinite Reynolds number, according to
$$\lim_{\tau_\eta\to 0}\tau_\eta^{2q}\E \left[\varphi_\tau^q\right]\build{\propto}_{\tau\to 0}^{}\left( \frac{T}{\tau}\right)^{\frac{\mu^l}{2}q(q-1)}e^{\frac{\mu^l}{2}q(q-1) \tilde g(0)}\int_{[0,1]^q}\prod_{i<j}^q\frac{1}{|s_i-s_j|^{\mu^l}}\prod_{i=1}^qds_i,$$
where again the exact expression of the multiplicative constant which is both scale and Reynolds number independent is given in \refp{eq:AsymptPhitauq}. The constant $\tilde{g}(0)$ has the same status as $g(0)$ in \refp{eq:MomEulPDMC} and is precisely defined alongside detailed computations in appendix \ref{Ann:CausMulAnalytic}.

\subsection{Differences between stochastic models of pseudo-dissipation and comparison with numerical flows}\label{Sec:DiffPDOUMulti}

As we have seen, modelling pseudo-dissipation as an exponentiated Ornstein-Uhlenbeck process, as done by \cite{GirPop90} and discussed in section \ref{Sec:GPDiss}, or as a multifractal measure, like discussed in \cite{Bor93} and considered by \cite{Sch03}, who proposed in the context of multiplicative chaos a causal dynamics based on heuristics arguments that we defined rigorously in section \ref{Sec:MultiDiss}, are very different matters. First of all, the first model is not consistent with the multifractal picture, which asks for (i) moments of $\varphi$ to behave as power laws of the Reynolds number, and (ii) moments of $\varphi_\tau$ to behave, in the asymptotic limit of infinite Reynolds number, as a power laws of the coarse-graining scale $\tau$. We have seen that indeed the second model, based on the multiplicative chaos, results in such statistical behaviour.

One may wonder whether these two different statistical behaviours could be tested against experiments and/or numerical simulations. Testing the behaviour of high order moments can be difficult since the different scalings are rather similar, and moreover, statistical convergence may be an issue. Instead, we could check the behaviour of the correlation function of the logarithm of pseudo-dissipation. Indeed, these two stochastic models have different predictions for the correlation of $\log \varphi$ along trajectories. On the one hand, the model of \cite{GirPop90} predicts that $\log \varphi$ is an Ornstein-Uhlenbeck process and so $\E \left[ \log \varphi(t) \log \varphi(t+\tau)\right]$ should decrease exponentially in time (over the integral time scale $T$).  On the other hand, multifractal modelling  asks for a logarithmic decrease over the time scale $T$. The analysis of \cite{HuaSch14}, based on Lagrangian trajectories extracted from a numerical simulation among the highest Reynolds numbers available from modern computers (with Taylor based Reynolds number on the order of 400) shows that data, and more precisely the correlation function of $\log \varphi$, are consistent both with exponential and logarithmic decreases. In other words, even at the highest accessible Reynolds numbers it is very difficult to make a distinction between the two models, both being able to reproduce in a quantitative way numerical data. In the following, we will prefer to work with the multifractal model, because (i) it has a proper behaviour in the asymptotic limit of infinite Reynolds numbers and (ii) the free parameter of the models, $\mu^l$, which appears in both models (see \refp{eq:Choiceabar} for the first model, and directly in \refp{eq:MCPDLag} for the second model) has a clear meaning only in a multifractal framework, where it is denominated the intermittency coefficient.

%%%%%%%%%%%%%%%%%%
%%  NEW SECTION %%
%%%%%%%%%%%%%%%%%%

\section{Stochastic dynamics for the velocity gradient tensor with multifractal properties}\label{Sec:InclMCRFD}

Let us now turn back to the construction of a stochastic model of the velocity gradient tensor $\tA$. To do so, let us first recall \refp{eq:SDE}, the general definition of a stochastic dynamics for $\tA$ along a Lagrangian trajectory, namely
\begin{equation}\label{eq:SDE2}
dA_{ij}(t) = V_{ij}(t)dt +D_{ijkl}(t) W_{kl}(dt),
\end{equation}
where $V_{ij}$ and $D_{ijkl}$ are respectively called the drift and diffusion terms, and $W_{ij}$ are the components of a tensorial white noise $\tW$, i.e. uncorrelated in time and with independent elements (not to be confused with the scalar noise, denoted simply by a serifed letter $W$). At a fixed time $t$, they are a priori causal functionals of the tensorial functions $\tA(s)_{s\in(-\infty,t]}$ and the tensorial Wiener process $\tW(s)_{s\in(-\infty,t)}$. Following the approach of \cite{GirPop90}, we want to determine $V_{ij}$ and $D_{ijkl}$ in order to (i) include some of the crucial physics of the Navier-Stokes equations \refp{eq:NSA} that we are able to close in terms of $\tA(t)$ and (ii) impose for the respective modelled pseudo-dissipation $\varphi=\tr(\tA\tA^\top)$ the multifractal structure described in section \ref{Sec:MultiDiss}.

Let us first remark that to constrain the dynamics of $\tA$ while imposing on one of its contractions $\varphi$ a certain precise statistical behaviour requires some approximations. Indeed, constraining $\varphi$, which is a scalar, does not determine uniquely $\tA$, which is a tensor, since several degrees of freedom are lost in the way. Similarly, we cannot at this stage start from \refp{eq:SDE2} and impose exactly for the respective pseudo-dissipation to be a multifractal process as described in section \ref{Sec:MultiDiss}. The reason is that this multifractal process is defined with a simple scalar white noise $W$ (see \refp{eq:MCPDLag} and \refp{eq:XText}), whereas the dynamics of $\tA$ is defined with a tensorial white noise $\tW$ \refp{eq:SDE2}. In the sequel, we will follow the ideas developed in \cite{GirPop90} to propose approximations that are consistent with underlying isotropic conditions and perform a mean-square estimation to determine the diffusion terms.

\subsection{General formulation and isotropic conditions.}

We consider \refp{eq:SDE2} as a modelled stochastic evolution of the velocity gradient tensor $\tA$. We are also interested in the respective dynamics of the pseudo-dissipation $\varphi=\tr (\tA\tA^\top)$ which is obtained from \refp{eq:SDE2} after applying the chain rule (Ito's lemma). Doing so, we obtain
\begin{equation}\label{eq:SDP1}
d\varphi= d\tr (\tA\tA^\top)
= \left[2A_{ij}V_{ij}+D_{ijkl}D_{ijkl}\right]dt + 2A_{ij}D_{ijkl}W_{kl}(dt).
\end{equation}
Assume now that the diffusion term $D_{ijkl}$ has a isotropic form, namely
\begin{equation}\label{eq:IsotropicD}
D_{ijkl} = a\delta_{ij}\delta_{kl}+b\delta_{ik}\delta_{jl}+c\delta_{il}\delta_{kj},
\end{equation}
where $a$, $b$, $c$ are unknown and, in general, scalar functionals of $\tA$ and time. The incompressibility condition imposes
\begin{equation}\label{eq:Incompabc}
3a+b+c=0.
\end{equation}
Computing the \textit{spurious} drift term $D_{ijkl}D_{ijkl}$ entering the dynamics of $\varphi$ \refp{eq:SDP1} and using the incompressibility condition $\tr (\tA)=A_{ii}=0$ results in the following dynamics for $\varphi$
\begin{equation}\label{eq:SDP2}
d\varphi= d\tr (\tA\tA^\top)
= \left[2A_{ij}V_{ij}+9(a^2+b^2+c^2)+6(ab+ac+bc)\right]dt + 2(bA_{kl}+cA_{lk})W_{kl}(dt).
\end{equation}

\subsection{Mean-square estimation of the diffusion term.}\label{Sec:MSDiff}

We want to impose a dynamics for $\varphi$ \refp{eq:SDP2} as close as possible to the one of a multiplicative chaos, derived in appendix \ref{Ann:CausMulAnalytic} and expressed in \refp{eq:sdePhi}, which we recall for convenience
\begin{align}\label{eq:sdePhiText}
d\varphi_{\tau_\eta} &= \varphi_{\tau_\eta}(t)\left[-\frac{1}{T} \left(\log [\tau_\eta^2\varphi_{\tau_\eta}(t)] +\frac{\mu^l}{2}\E \left[\left(X_{\tau_\eta}\right)^2\right]\right)+\sqrt{\mu^l}\beta_{\tau_\eta}(t) +\frac{\mu^l}{2{\tau_\eta}}\right]dt\notag\\
& + \sqrt{\frac{\mu^l}{\tau_\eta}}\varphi_{\tau_\eta}(t)W(dt).
\end{align}
This can be done at the cost of an approximation since, as explained, the true dynamics of $\varphi$ \refp{eq:SDP2}, comprising a tensorial white noise $\tW$, involves more degrees of freedom than the multiplicative chaos one \refp{eq:sdePhiText}, where only a scalar white noise $W$ appears.

We follow the approximation of \cite{GirPop90} that consists in adopting a mean-square approximation of the diffusion term. We thus take the diffusion term from \refp{eq:SDP1} such that its variance conditioned to $\tA$ equals the variance of the diffusion term from \refp{eq:sdePhiText} conditioned to $\varphi_{\tau_\eta} = \varphi = A_{ij}A_{ij}$. Since we are dealing with random distributions $\tW$ and $W$, the meaning of their square is a priori not clear. Nonetheless, this can be clarified while considering the square of their integral over a finite range and then taking a limit as depicted in \cite{GirPop90}.
Doing so, we obtain, for any time $t$,
 \begin{align}\label{eq:EquMSE}
4\left[(b^2+c^2)\varphi+2bc \,\tr(\tA^2)\right]= \frac{\mu^l}{\tau_{\eta}}\varphi^2.
\end{align}
As in \cite{GirPop90}, we remark that the equality in \refp{eq:EquMSE} provides an underdetermined problem in relation to the free parameters $b$ and $c$ (one equation for two unknowns). To this regard, we assume that the dynamics of $\varphi=\tr(\tA\tA^\top)$ only depends on itself, and not on the contraction $\tr(\tA^2)$.  Using furthermore the incompressibility condition \refp{eq:Incompabc} we arrive at
\begin{equation}\label{eq:Diffabc}
a=-\frac{1}{6}\sqrt{\frac{\mu^l\varphi}{\tau_{\eta}}}\mbox{ , }b=\frac{1}{2}\sqrt{\frac{\mu^l\varphi} {\tau_{\eta}}} \mbox{ and }c=0.
\end{equation}
It would be interesting to study the dependence of the model on the choice $c\ne 0$, particularly if we want to build up separately dynamics for dissipation $\tr (\tS^2)$ and enstrophy $-\tr (\tO^2)$. We keep these developments for further investigations. The dynamics of $\varphi$ \refp{eq:SDP2} becomes
\begin{equation}\label{eq:SDP3}
d\varphi= d\tr (\tA\tA^\top)
= \left[2A_{ij}V_{ij}+2\frac{\mu^l}{\tau_{\eta}}\varphi\right]dt + \sqrt{\frac{\mu^l\varphi} {\tau_{\eta}}} A_{kl}W_{kl}(dt).
\end{equation}
We can see that this mean-square procedure suggests the replacement of the scalar white noise $W$ in the dynamics of $\varphi_{\tau_\eta}$ \refp{eq:sdePhiText} by the scalar noise $A_{kl}W_{kl}/\sqrt{\varphi}$ that emerges from the $\varphi$ dynamics \refp{eq:SDP3}.

\subsection{Determination of the drift term and introduction of the Recent Fluid Deformation closure}
\label{Sec:DetermDriftRFD}
Analogously to \cite{GirPop90}, we would like to impose on $\varphi$ \refp{eq:SDP3} some dynamics as close as possible to the one fulfilled by the multiplicative chaos \refp{eq:sdePhiText}. To do so, we need to provide an expression for the tensor $\tV$ of the drift part of \refp{eq:SDP3}. This tensor is the only one able to give a flavour of the Navier-Stokes equation \refp{eq:NSA}, and should be closed in terms of $\tA$. The very first idea would be to take into account only the self-stretching term $\tV=\tV^{\mbox{\tiny{RE}}}$ \refp{eq:REDeterm}. This would be done disregarding all the interesting physics of the joint action of the pressure Hessian and viscous term which compose the true dynamics of $\tA$ \refp{eq:NSA}. So instead, we will adopt the more sophisticated closure $\tV=\tV^{\mbox{\tiny{RFD}}}$ \refp{eq:RFDDeterm} given by the Recent Fluid Deformation (RFD) approximation and discussed in \cite{CheMen06,CheMen08,Men11}. Although it reproduces some of the physics governed by the pressure and viscosity terms, this closure is not consistent with the constraint we want to impose for the pseudo-dissipation \refp{eq:sdePhiText} when inserted in \refp{eq:SDP3}. To make it so, we will furthermore consider an additional damping term proportional to $\tA$ and show that the associated multiplicative factor can be chosen such that the resulting dynamics for $\varphi$ is similar, in a certain sense, to the one fulfilled by the multiplicative chaos \refp{eq:sdePhiText}. This is similar to the procedure made by \cite{GirPop90}. Accordingly, we choose for the drift tensorial term $\tV$ \refp{eq:SDE2} the following model
\begin{equation}\label{eq:DriftV}
\tV(t) = \tV^{\mbox{\tiny{RFD}}}(t)+f(t)\tA(t),
\end{equation}
where the multiplicative factor $f(t)$ is chosen so that the respective drift term of the dynamics of $\varphi$ is as close as possible to the one of the multiplicative chaos \refp{eq:sdePhiText}. With this aim, and to find an expression for $f(t)$, we must identify the drift parts of the dynamics of $\varphi_{\tau_\eta}$ \refp{eq:sdePhiText} with the those of the $\varphi$ dynamics \refp{eq:SDP3}.

As we have seen in section \ref{Sec:MSDiff}, the application of a mean-square procedure to the diffusion term entering in the dynamics of $\varphi$ \refp{eq:SDP3} suggests to replace the scalar noise $W$ by $A_{kl}W_{kl}/\sqrt{\varphi}$. We will follow this suggestion and replace accordingly the fractional noise $\beta_{\tau_\eta}$ entering \refp{eq:sdePhiText} and defined in \refp{eq:betaText}, built from the scalar noise $W$, by its respective version $\hat{\beta}_{\tau_\eta}$ built on the noise $A_{kl}W_{kl}\sqrt{\varphi}$. This new noise reads
\begin{equation}\label{eq:betahat}
\hat{\beta}_{\tau_\eta}(t) = -\frac{1}{2} \int_{s=-\infty}^t \frac{1}{(t-s+{\tau_\eta})^{3/2}} \frac{A_{ij}(s)}{\sqrt{\varphi(s)}}W_{ij}(ds).
\end{equation}
Several reasons pushed us to consider this noise $\hat{\beta}_{\tau_\eta}$. First of all, as mentioned, the mean square procedure used in section \ref{Sec:MSDiff} to determine the diffusion term $D_{ijkl}$ suggests the replacement of $W$ by $A_{kl}W_{kl}/\sqrt{\varphi}$. One could try to replace the scalar noise $W$ by a another scalar such as $\tr(\tW)$, but in doing so one loses the correlation between $\hat{\beta}_{\tau_\eta}$ and the random term $A_{kl}W_{kl}$ from the dynamics of $\varphi$ \refp{eq:SDP3}. This correlation, as shown in the appendix \ref{Ann:MultiChaosDiss}, is crucial in order to achieve multifractal statistics. One could then wonder that the statistics of $\hat{\beta}_{\tau_\eta}$ might be completely different from the initial Gaussian noise $\beta_{\tau_\eta}$ that forms the multiplicative chaos dynamics \refp{eq:sdePhiText} and leads to exact multifractal properties of the positive field $\varphi_{\tau_\eta}$. It is to be checked numerically, a posteriori, that their imprints on the final dynamics have similar consequences. And indeed, as we verify in section \ref{sec:NumRes}, dedicated to numerical results, the model we are constructing with the noise $\hat\beta_{\tau_\eta}$ culminate in a pseudo-dissipation whose statistics is remarkably similar to the exact multifractal model $\varphi_{\tau_\eta}$ from \refp{eq:sdePhiText}. This being said, it is possible to verify numerically that the variances of $\hat{\beta}_{\tau_\eta}$ and $\beta_{\tau_\eta}$ are actually the same, i.e.
\begin{equation}\label{eq:EqHatnoHat}
\E \left[\hat{\beta}^2_{\tau_\eta}\right] \approx \E \left[\beta^2_{\tau_\eta}\right].
\end{equation}
Thus, as far as the variance is concerned, the noises $\hat{\beta}_{\tau_\eta}$ \refp{eq:betahat} and $\beta_{\tau_\eta}$ \refp{eq:betaText} behave similarly. Higher order moments will be different though, since $\beta_{\tau_\eta}$ is a Gaussian process and $\hat{\beta}_{\tau_\eta}$ is not.

Again analogously to \cite{GirPop90}, we equate the drift term of the multifractal process \refp{eq:sdePhiText} with the drift term of pseudo-dissipation \refp{eq:SDP3} using as a model for $\tV$ the sum of the RFD closure's drift term and a (non-linear) damping term \refp{eq:DriftV}, while replacing the noise $\beta_{\tau_\eta}$ by the new noise $\hat{\beta}_{\tau_\eta}$, to deduce an expression for the functional $f(t)$
\begin{equation}\label{eq:ExpressfA}
f(t) = -\frac{1}{2T} \left(\log [\tau_\eta^2\varphi(t)]+\frac{\mu^l}{2}\E \left[\left(X_{\tau_\eta}\right)^2\right]\right)+\frac{\sqrt{\mu^l}}{2}\hat{\beta}_{\tau_\eta}(t) -\frac{3\mu^l}{4{\tau_\eta}} - \frac{A_{ij}(t)V_{ij}^{\mbox{\tiny{RFD}}}(t)}{\varphi(t)}.
\end{equation}
Let us note that the quantity $\E \left[\left(X_{\tau_\eta}\right)^2\right]$, a function of $\tau_\eta/T$ only (see \refp{eq:ResVar}), already present in the dynamics of $\varphi_{\tau_\eta}$ \refp{eq:sdePhiText}, also appears in \refp{eq:ExpressfA}.

\subsection{Assembling the model}

By inserting the various terms discussed above in the dynamics of $\tA$ \refp{eq:SDE2}, we forge a closed dynamics along a trajectory. For clarity and convenience, we recall here all the pieces gathered to build the model. For the drift term \refp{eq:DriftV}, $\tV=\tV^{\mbox{\tiny{RFD}}} + f(t)\tA$, there are two contributions. The RFD part is given by \refp{eq:RFDDeterm}, that is,
\begin{equation}
\tV^{\mbox{\tiny{RFD}}} = -\tA^2+\frac{\tr(\tA^2)}{\tr(\tC_{\tau_\eta}^{-1})}\tC_{\tau_\eta}^{-1} -\frac{\tr(\tC_{\tau_\eta}^{-1})}{3T}\tA,
\end{equation}
being $\tC_{\tau_\eta}$ the short-time Cauchy-Green tensor \refp{eq:CGtaueta} and $\tau_\eta$ and $T$ the Kolmogorov and integral time scales respectively. The damping part is formed with the scalar function $f(t)$ derived just above in \refp{eq:ExpressfA}, which comprises the intermittency coefficient $\mu^l$, the quantity $\E \left[\left(X_{\tau_\eta}\right)^2\right]$, which is function of $\tau_\eta/T$ only, and justified in \refp{eq:ResVar}, that reads
\begin{equation}\label{eq:ResVarRepeat}
\E \left[\left(X_{\tau_\eta}\right)^2\right] =\int_{\mathbb R^+}\frac{e^{-\frac{h}{T}}}{h+\tau_\eta+\sqrt{\tau_\eta(h+\tau_\eta)}}dh,
\end{equation}
and can be eventually expressed with special functions, as given by a symbolic calculation software. A modification to the regularized fractional Gaussian noise of vanishing Hurst exponent \refp{eq:betaText} denoted as $\hat\beta_{\tau_\eta}$ \refp{eq:betahat}, which reads
\begin{equation}\label{eq:betahat2}
\hat{\beta}_{\tau_\eta}(t) = -\frac{1}{2} \int_{s=-\infty}^t \frac{1}{(t-s+{\tau_\eta})^{3/2}} \frac{A_{ij}(s)}{\sqrt{\varphi(s)}}W_{ij}(ds),
\end{equation}
enters also in the definition of the scalar function $f(t)$. Together with the isotropic diffusion term $D_{ijkl}$ (equations \ref{eq:IsotropicD} and \ref{eq:Diffabc}), constructed by a mean-square estimation procedure (section \ref{Sec:MSDiff}), one finally arrives at
\begin{equation}\label{eq:DynAFinal}
dA_{ij}(t) = \left[V^{\mbox{\tiny{RFD}}}_{ij}(t)+f(t)A_{ij}(t)\right]dt + \frac{1}{2}\sqrt{\frac{\mu^l\varphi(t)}{\tau_\eta}}\left[W_{ij}(dt)-\frac{1}{3}\tr[\tW(dt)]\delta_{ij}\right].
\end{equation}
This proposed dynamics is closed. It yields, starting for instance at time $t=0$, the time evolution of $\tA(t)_{t\ge 0}$ once some history $\tA(t)_{t< 0}$ is given. This boundary condition is required by the modification of the fractional Gaussian noise $\hat{\beta}_{\tau_\eta}$ \refp{eq:betahat2} that seeks in the past to set its current value. In this sense the dynamics proposed in \refp{eq:DynAFinal} is non-Markovian. We expect this noise to decorrelate as fast as $1/t^2$ because its Gaussian version $\beta_{\tau_\eta}$ does decorrelate this way (see developments in appendix \ref{Ann:MultiChaosDiss}). So in practice, the noise $\hat{\beta}_{\tau_\eta}(t)$ \refp{eq:betahat2} is well approximated by a truncation over a finite interval, say for instance $t\in[t-T,t]$, and this approximation is more accurate as the small time scale $\tau_\eta$ gets smaller compared to $T$. %This being said, the model \refp{eq:DynAFinal} for $\tA$ depends moreover on the large time scale $T$, the small time scale $\tau_\eta$, and the intermittency coefficient of the Lagrangian frame $\mu^l$.

%%%%%%%%%%%%%%%%%%
%%  NEW SECTION %%
%%%%%%%%%%%%%%%%%%

\section{Numerical procedure}\label{sec:Num}

This section is devoted to the numerical integration of the proposed new dynamics for the velocity gradient tensor $\tA$ along a trajectory \refp{eq:DynAFinal}. For this purpose, we need a numerical approximation for the noise $\hat{\beta}_{\tau_\eta}[t]$ \refp{eq:betahat} which composes the function $f(t)$ \refp{eq:ExpressfA}, as well as its boundary condition. Denoting as $\Delta t$ a time marching step, the tensor $\tW[dt]$ is such that each element $W_{ij}[dt]$ at time $t$ is an independent realization of a zero-mean Gaussian random variable with variance $\Delta t$.

Over a discrete set of time and starting, say, at time $t=0$, we will use henceforth the following numerical approximation of the noise $\hat{\beta}_{\tau_\eta}$ \refp{eq:betahat}:
\begin{equation}\label{eq:betahatNA}
\hat{\beta}_{\tau_\eta}[t] \approx -\frac{1}{2} \sum_{s=t-T}^t \frac{1}{(t-s+{\tau_\eta})^{3/2}}  \frac{A_{ij}[s]}{\sqrt{\varphi[s]}}W_{ij}[ds].
\end{equation}
This approximation, based on a truncation of the time integration, is, as mentioned in the end of section \ref{Sec:DetermDriftRFD}, more and more realistic as $\tau_\eta/T\to 0$. A numerical study devoted to the dependence of the approximation of the fractional Gaussian noise \refp{eq:betahat} by its truncated version \refp{eq:betahatNA} remains to be properly done, since this is the most demanding step of the numerical integration of the $\tA$ dynamics. Nevertheless, we have performed simulations (data not shown) using a truncation over a shorter range $[t-T/2,t]$ and no relevant quantitative differences were observed on the statistical results, at least in the investigated range of $\tau_\eta$ values. To make sense of the discretization used in \refp{eq:betahatNA}, we also need to choose $\Delta t$ small enough compared to $\tau_\eta$ to properly resolve the smooth kernel $(t-s+{\tau_\eta})^{-3/2}$ when $s\to t^-$.

Having set a numerical approximation of the noise $\hat{\beta}_{\tau_\eta}$ \refp{eq:betahatNA}, we need now to give initial conditions. In the following numerical simulations, we will start at time $t=0$, with $\tA[0]=\tNo-\tr(\tNo)\tI/3$, where $\tNo$ is a $3\times 3$ tensor such that elements are independent and normally distributed with zero average and unit variance. As boundary conditions, we simply take $\tA[t]_{-T\le t\le 0}=\tA[0]$. The precise forms of the initial and boundary conditions are not important (as long at they are trace-free) since we expect to reach a stationary regime independent of them at large time $t$. Finally, we use for time marching a simple Euler approximation, which is a delicate matter and deserve some comments. Even the multiplicative chaos dynamics \refp{eq:sdePhiText} with the fractional Gaussian noise \refp{eq:betaText} exhibits a rather slow statistical convergence with respect to the ensemble size, as already observed for the Eulerian case in \cite{PerGar16}, where many realizations of high resolution 3d cubes were needed to assert their conclusions. Thus very large ensembles will be necessary to attain smooth statistics. With the Euler method, small time steps are required to reduce the numerical error associated with time integration, meaning that a fairly big amount of data will be needed, in particular for large Reynolds numbers, but this may translate into a substantial error accumulation towards the end of the integration. Unfortunately, for the proposed dynamics \refp{eq:DynAFinal}, the highly non-linear and non-Markovian character, together with the multiplicative nature of the noise and the strong correlations between the terms, make the implementation of higher order time integration methods a challenging problem. All things considered, we employ the Euler method but instead of producing very long trajectories we choose to generate a large ensemble of moderately shorter trajectories (always respecting an initial transient time). Also, for the larger Reynolds numbers, runs with varying $\Delta t$ revealed decisive to determine if discrepancies among the statistical behaviours of $\varphi$ built from \refp{eq:DynAFinal} and $\varphi_\tau$ from \refp{eq:sdePhiText} were caused by numerical errors or fundamental differences between the processes.

%%%%%%%%%%%%%%%%%%
%%  NEW SECTION %%
%%%%%%%%%%%%%%%%%%

\section{Numerical results}\label{sec:NumRes}

\subsection{Parameters of the simulations}\label{sec:ParamNum}

In the following simulations we rescale time by $T$ and take $T=1$, with no loss of generality. As for the intermittency coefficient in the Lagrangian framework $\mu^l$, we use the recently estimated value on high Reynolds number numerical turbulent data of \cite{HuaSch14}, who found $\mu^l=0.3$.
Notice that this is consistent with the commonly accepted value of the intermittency coefficient in the Eulerian framework $\mu^e=0.2$, as it may be checked with the phenomenological theory of \cite{Bor93}, which relates $\mu^l$ and $\mu^e$ through a non-linear relation (relating more precisely the respective singular spectra, using the vocabulary of the multifractal formalism). If one neglects the fluctuating nature of the spatial and temporal dissipative scales, an approximative linear relation can be drawn in a straightforward way: we observe that the moments of the pseudo-dissipation field $\varphi$ can be expressed in the Eulerian \refp{eq:MomEulPDMC} and Lagrangian \refp{eq:PredMomqLag} frameworks, and, recalling that $\eta/L \propto \mathcal R_e^{-3/4}$ and $\tau_\eta/T \propto \mathcal R_e^{-1/2}$, one is led to $3\mu^e=2\mu^l$ by equating their Reynolds number dependence.

The positive quantity $\E [(X_{\tau_\eta})^2]$ contained in the dynamics of $\tA$ \refp{eq:DynAFinal} is given explicitly as an integral in \refp{eq:ResVar}. This integral could be numerically evaluated, but it may be analytically expressed in terms of error and generalized hypergeometric functions, which in turn must be numerically evaluated as well, but whose handling by popular math softwares can be quite efficient and controllable. We follow this path to compute $\E [\left(X_{\tau_\eta}\right)^2]$ for the various values of $\tau_\eta/T$ considered.

Using $\Delta t=2\times 10^{-3}\tau_\eta$, or $\Delta t=10^{-3}\tau_\eta$ for higher $\mathcal R_e$, we numerically integrate numerous realizations of the $\tA$ dynamics over time intervals of dozens or hundreds of $T$, such that at least a total time of $3\times 10^3\, T$ is covered. We observe then, for the explored range $2.48\times 10^{-3} \le \tau_\eta/T \le 2.64 \times 10^{-1}$, or equivalently $-6.0 \le \log(\tau_\eta/T) \le -1.33$, the stationary statistics of $\tA$ and its contraction $\varphi = A_{ij}A_{ij}$.

It is worth mentioning that extensive tests were successfully done in order to establish not only the stationarity of the generated processes but also its independence of the initial conditions. In this perspective, we checked (data not shown) that an ensemble of highly anisotropic initial conditions indeed relaxes to an isotropic state, at least with the same level of success as in subsection \ref{sec:StatVarPhi}.

\subsection{Variance and covariance of the velocity gradients, and the Reynolds number dependence}

\subsubsection{Variance}

\begin{figure}
\begin{center}
\input{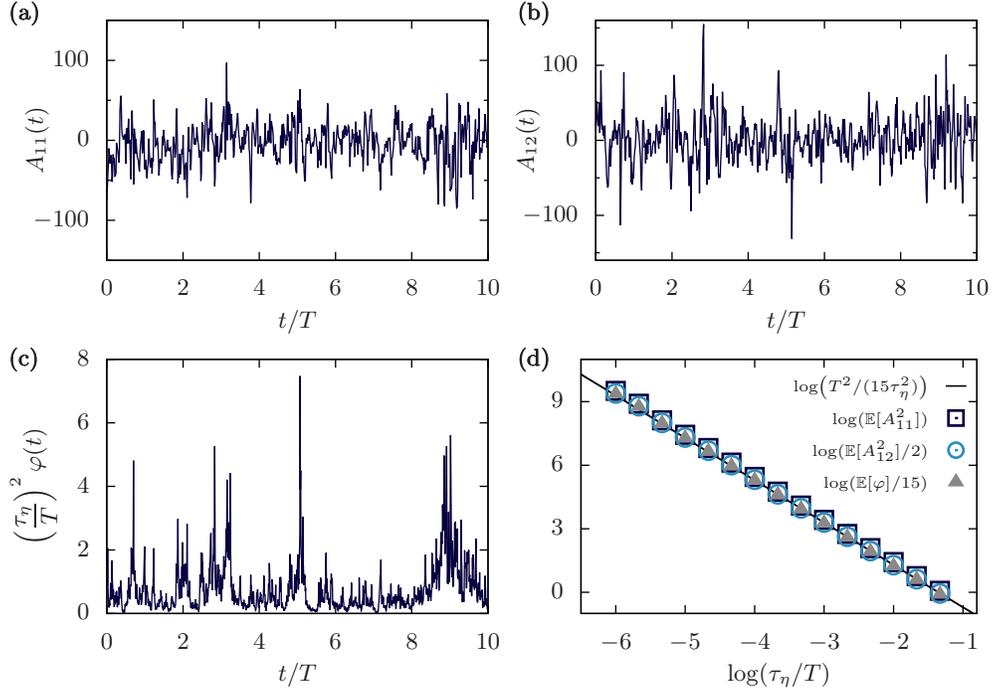}
\end{center}
\caption{\label{fig:VarTauEta}  In arbitrary units, samples of (a) a diagonal element $A_{11}$, (b) an off-diagonal element $A_{12}$ and (c) the implied pseudo-dissipation $\varphi$ along a trajectory for $\log(\tau_\eta/T)=4.67$ showing characteristic erratic and intermittent fluctuations. (d) Variance of diagonal (i.e. $A_{11}$, $\square$) and off-diagonal (i.e. $A_{12}$, $\modcirc$) components and expectation value of pseudo-dissipation $\varphi$ ($\blacktriangle$) as a function of the ratio of the two characteristic scales $\tau_\eta/T$. The solid line shows in this representation the power law $\frac{1}{15}\left(\frac{T}{\tau_\eta}\right)^2$}
\end{figure}

For illustrative purposes, let us first begin by showing some temporal profiles of a diagonal element $A_{11}$ (figure \ref{fig:VarTauEta}a), an off-diagonal element $A_{12}$ (figure \ref{fig:VarTauEta}b) and the related pseudo-dissipation $\varphi$ (figure \ref{fig:VarTauEta}c) along a Lagrangian trajectory, at a given Reynolds number quantified by $\log(\tau_\eta/T)=4.67$. We observe an erratic behaviour of gradients that is expected in a highly turbulent flow. We may also notice the non-symmetrical behaviour of the diagonal element $A_{11}$ fluctuations (figure \ref{fig:VarTauEta}a), reminiscent of the skewness phenomenon, and the highly intermittent $\varphi$ fluctuations. We invite the reader to take a look at section \ref{eq:NumStudyHO} for a more in-depth study of the non-Gaussian nature of these fluctuations.

The second behaviour we would like to check is whether the proposed stochastic model \refp{eq:DynAFinal} is consistent with basic Kolmogorov phenomenology, such as the Reynolds number (or $\tau_\eta/T$) dependence of the variance of the elements of $\tA$. In particular, we would like the average pseudo-dissipation $\E(\varphi)=\E[\tr(\tA\tA^\top)]$ to be proportional to $(T/\tau_{\eta})^2$, according to \refp{eq:KolPredPD}. Let us notice that we expect that since we are somehow putting this property by hands. Indeed, we use as a constraint for building the dynamics of $\tA$ \refp{eq:DynAFinal} that it fulfils as close as possible the dynamics of a multiplicative chaos $\varphi_\eta$ \refp{eq:sdePhiText}, which is such that  $\E(\varphi_\eta)=1/\tau_{\eta}^2$. It remains to check numerically that this behaviour is actually observed for $\varphi$. More precisely, we would like to check whether, as required by isotropy \citep[see for instance][]{Pop00},
\begin{equation}\label{eq:VerNumVarEle}
\E (A_{11}^2) = \frac{1}{2}\E (A_{12}^2) = \frac{1}{15}\E (\varphi) =\frac{1}{15}\left(\frac{T}{\tau_\eta}\right)^2.
\end{equation}
We represent in figure \ref{fig:VarTauEta}(d) in a logarithmic scale the behaviour of the variances of components $A_{11}$ and $A_{12}$ and the average of $\varphi$ as a function of the ratio $\tau_\eta/T$. We see that the isotropic relations \refp{eq:VerNumVarEle} are verified on our numerical simulations of the process $\tA$ \ref{eq:DynAFinal} for a wide interval of $\tau_\eta/T$. This shows that, in this sense, statistics of $\tA$ are isotropic and behave in a consistent way with $\tau_\eta/T$.

\subsubsection{Autocovariance}

\begin{figure}
\begin{center}
\input{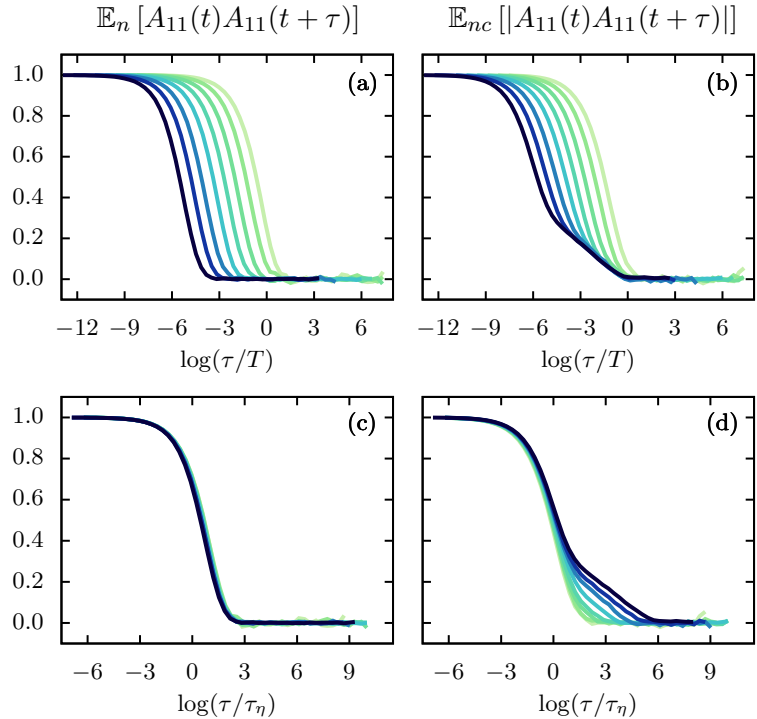}
\end{center}
\caption{\label{fig:CorrA} Autocorrelation functions of the diagonal element $A_{11}$, as a function of the time lag. We represent in (a) and (c) the correlation function $\E_n\left[A_{11}(t)A_{11}(t+\tau)\right]$ (defined in \refp{eq:DefEn}), and in (b) and (d) the correlation functions of the absolute values $\E_{nc}[|A_{11}(t)A_{11}(t+\tau)|]$ (defined in \refp{eq:DefEnc}). We show data for $\log(\tau_\eta/T)= -1.33,-2.0,-2.67,-3.33,-4.0,-4.67,-5.33,-6.0$ (from lighter to darker). In (a) and (b) time lag is normalized by $T$, whereas in (c) and (d) by $\tau_\eta$.}
\end{figure}

As far as second order statistics are concerned, another important property that should be checked in the simulated trajectories of $\tA$ is the covariance of its elements. Kolmogorov's phenomenology \citep{Fri95} suggests that, along Lagrangian trajectories, elements of $\tA$ (such as $A_{11}$ or $A_{12}$) are correlated over the small time scale $\tau_\eta$. We remark that if indeed this short-time correlation is observed in the present stochastic model, it is not clear to us how to derive it directly from the dynamics given in \refp{eq:DynAFinal} \citep[see devoted works on this subject by][]{AfoMen10,YuMen10} since we are only imposing that pseudo-dissipation (which is a positive quantity) is correlated over the large time scale $T$. We will come back to this long-range correlation behaviour in the next section. As we are dealing with a highly nonlinear dynamics and not imposing in a rigorous and definitive way that the elements of $\tA$ should be correlated over the short time scale $\tau_\eta$, it deserves to be precisely quantified in our simulations. We represent in figure \ref{fig:CorrA} the estimation of the correlation function of the diagonal element $A_{11}$ as a function of (i) the rescaled time $\tau/T$ (figure \ref{fig:CorrA}(a)), and (ii) the rescaled time $\tau/\tau_{\eta}$ (figure \ref{fig:CorrA}(b)). Since the variance of $A_{11}$ increases as $\tau_{\eta}/T$ decreases, we focus on the normalized covariance, namely
\begin{equation}\label{eq:DefEn}
\E_n\left[A_{11}(t)A_{11}(t+\tau)\right] = \frac{1}{\E\left[A_{11}^2\right]}\E\left[A_{11}(t)A_{11}(t+\tau)\right],
\end{equation}
which is independent of $t$ by stationarity, and tends towards unity as $\tau\to 0$. Note also that we do not subtract the square of the average of the diagonal element since we have $\E(A_{11})=0$. A comparison between figures \ref{fig:CorrA}(a) and \ref{fig:CorrA}(c) shows clearly that the diagonal element $A_{11}$ is indeed correlated over the Kolmogorov's time scale $\tau_\eta$. Similar observations can be made on the off-diagonal element $A_{12}$ (data not shown).

As we have seen, the diagonal element $A_{11}$ is correlated over the small time scale $\tau_\eta$, as it should be. One may wonder what is the characteristic correlation time scale of the amplitude of the diagonal element $A_{11}$, or equivalently the absolute value of the diagonal element $|A_{11}|$. To do so, we define the following (normalized and centred) autocorrelation function
\begin{equation}\label{eq:DefEnc}
\E_{nc}\left[\Big|A_{11}(t)A_{11}(t+\tau)\Big|\right] = \frac{\E\left[\left(|A_{11}(t)|-\E[|A_{11}|]\right)\left(|A_{11}(t+\tau)|-\E[|A_{11}|]\right)\right]}{\E\left[\left(|A_{11}|-\E[|A_{11}|]\right)^2\right]},
\end{equation}
which, by stationarity, should be time independent.

We represent in figure \ref{fig:CorrA} the behaviour of this correlation function \refp{eq:DefEnc} both as a function of $\log(\tau/T)$ (figure \ref{fig:CorrA}(b)) and $\log(\tau/\tau_{\eta})$ (figure \ref{fig:CorrA}(d)). We indeed observe two type of behaviours. When the lag $\tau$ is rescaled by the large scale $T$, figure \ref{fig:CorrA}(b) reveals that the short time portion of the correlations, typically when $\tau$ is of the order of $\tau_\eta$, does not coincide, whereas for large times (i.e. when $\tau\gg \tau_\eta$) the correlations vanish around the large scale $T$. This analysis is confirmed in figure \ref{fig:CorrA}(d) where it is seen that the correlations superimpose over the short time scale $\tau_\eta$ but not over larger time lags.

This study shows that the (signed) diagonal elements decorrelate over the short time scale $\tau_\eta$, whereas their absolute values decorrelate over the large time scale $T$. This is a hallmark of intermittency: the sign of fluctuations decorrelates very fast, whereas the amplitude remains correlated over the large time scale $T$. This is consistent with saying that intermittency and high values of gradients appear as \textit{bursts}, that can be correlated over possibly large time scales \citep{Fri95}.

\subsection{Comparisons between the dynamics of $\varphi$ and $\varphi_{\tau_\eta}$} \label{sec:StatVarPhi}

\begin{figure}
\begin{center}
\input{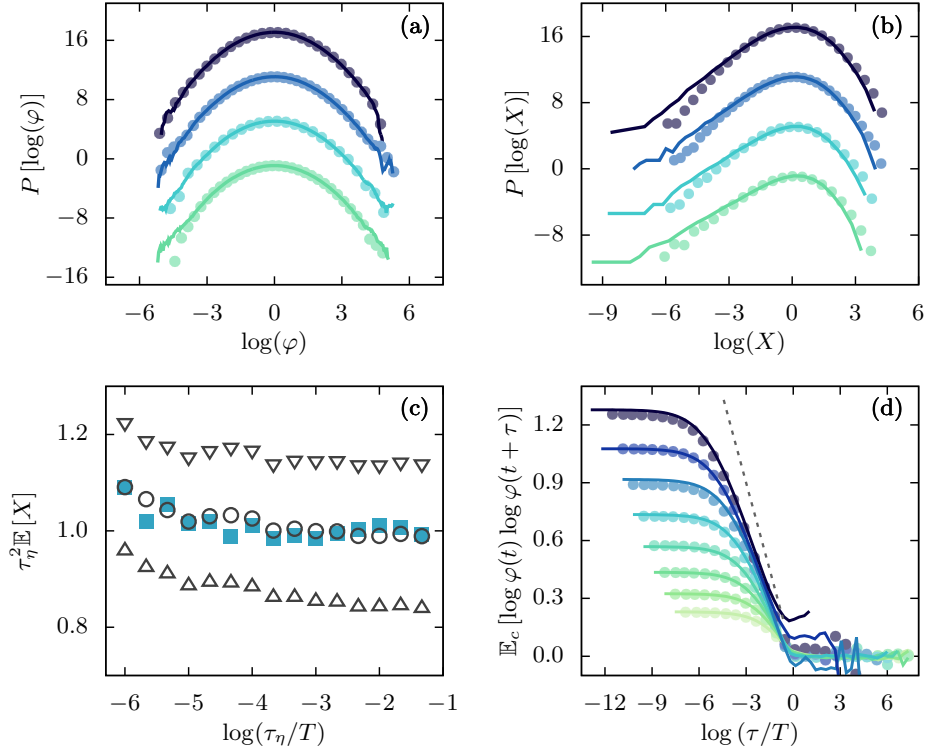}
\end{center}
\caption{\label{fig:StatMeasures} (a) Logarithmic representation of the resulting PDFs of $\log \varphi_{\tau_\eta}$ (lines) and $\log \varphi$ (points), centred at the origin, normalized to unit variance and vertically shifted arbitrarily for clarity. Values of $\log(\tau_\eta/T)$ range from -2.67 to -5.67 in unit steps (from lighter to darker). (b) PDFs of dissipation (lines) and enstrophy (points) for the same ratios $\tau_\eta/T$. (c) Averages $\E[\varphi]$ (\modcirc), $\E[\varphi_{\tau_\eta}]$ (\modsquare), $2\E[\tr(\tS\tS^\top)]$ ($\bigtriangledown$) and $2\E[\tr(\tO\tO^\top)]$ ($\bigtriangleup$), as functions of $\log(\tau_\eta/T)$. (d) Autocovariance $\E_c$ \refp{eq:DefEcMeas} as a function of time lag for $\log \varphi$ (lines) and $\log \varphi_{\tau_\eta}$ (circles). Values of $\log(\tau_\eta/T)=$ span from -1.33 to -6.0 (lighter to darker) just as in figure \ref{fig:CorrA}. The asymptotic logarithm behaviour \refp{eq:BehavEcPhiTauEta} fulfiled by $\varphi_{\tau_\eta}$ is also shown for comparison (dashed line).}
\end{figure}

In this section, we would like to study numerically the statistical properties of the pseudo-dissipation $\varphi= \tr(\tA\tA^\top)$ as obtained from the stochastic process \refp{eq:DynAFinal}, which are difficult to get analytically. We are also interested in comparing them to the ones given by the model $\varphi_{\tau_\eta}$ \refp{eq:sdePhiText}, that can be carried on analytically (appendix \ref{Ann:MultiChaosDiss}). The statistics of $\varphi$ and $\varphi_{\tau_\eta}$ are indeed expected to be very similar, since their dynamics are themselves driven by similar equations. To justify this, compare the dynamics of $\varphi_{\tau_\eta}$ \refp{eq:sdePhiText} to that of $\varphi$, which can be obtained while inserting \refp{eq:ExpressfA} in the dynamics of $\tA$ \refp{eq:DynAFinal} and applying the chain rule, which reads
\begin{align}\label{eq:sdePhiAFinal}
d\varphi &= \varphi(t)\left[-\frac{1}{T} \left(\log [\tau_\eta^2\varphi(t)] +\frac{\mu^l}{2}\E \left[\left(X_{\tau_\eta}\right)^2\right]\right)+\sqrt{\mu^l}\hat{\beta}_{\tau_\eta}(t) +\frac{\mu^l}{2{\tau_\eta}}\right]dt\notag\\
& + \sqrt{\frac{\mu^l}{\tau_\eta}}\sqrt{\varphi(t)}A_{kl}W_{kl}(dt).
\end{align}
We see that \refp{eq:sdePhiText} and \refp{eq:sdePhiAFinal} are different for two reasons: (i) the noise $\beta_{\tau_\eta}$ \refp{eq:betaText} is replaced by the noise $\hat{\beta}_{\tau_\eta}$ \refp{eq:EqHatnoHat} and (ii) the noise $\varphi_{\tau_\eta}W$ is replaced by $\sqrt{\varphi(t)}A_{kl}W_{kl}$.

Continuing our analysis, we represent in figure \ref{fig:StatMeasures}(a) the estimation of the PDFs of the realizations of the random variables $\log \varphi_{\tau_\eta}$ and $\log \varphi$, once centred (we subtract their respective means) and normalized (we divide them by their respective standard deviations) for different values of the ratio $\tau_\eta/T$. We recall that $\varphi_{\tau_\eta}$ is obtained while exponentiating $X_{\tau_\eta}$, accordingly with \refp{eq:MCPDLag}, which is itself given by a linear stochastic differential equation \refp{eq:sdeXText}. We see that, as expected, the statistics of $\log \varphi_{\tau_\eta}$ is indeed normal for any $\tau_\eta/T$. As for $\log \varphi$, its PDFs closely follow those of $\log \varphi_{\tau_\eta}$, and hence their fluctuations are indeed observed very similar.

As also studied by \cite{YeuPop89,Pop90,HuaSch14}, we reproduce in figure \ref{fig:StatMeasures}(b) a similar estimation of the PDFs of two other quantities derived from $\tA$, namely dissipation $\tr(\tS\tS^\top)$ and enstrophy $\tr(\tO\tO^\top)$, being $\tS$ and $\tO$ respectively the symmetric and antisymmetric parts of the decomposition of $\tA$. We see that whereas the statistics of $\varphi$ is very close to the one of a lognormal process, those of dissipation and enstrophy differ significantly from such a process. This fact holds also true in direct numerical simulations (DNS) of turbulence as shown in \cite{YeuPop89,Pop90,HuaSch14}.

Let us now focus on the averages of these fields. We represent in figure \ref{fig:StatMeasures}(c) the average values of the pseudo-dissipation fields $\varphi$ and $\varphi_{\tau_\eta}$. As expected, since this is imposed by hands, it is observed that $\tau_\eta^2\E [\varphi_{\tau_\eta}]\approx 1$ for all studied values of $\tau_\eta/T$. Up to statistical fluctuations, we also observe that $\E [\varphi]\approx\E [\varphi_{\tau_\eta}]$ for all $\tau_\eta/T$. Here, numerous tests with smaller time steps and bigger ensembles were necessary to establish this result, since systematic differences may appear at high Reynolds numbers if statistical convergence is not properly attained. In fact, the mild increases in the averages one observes in figure \ref{fig:StatMeasures}(c) for the smaller ratios $\tau/T$ (higher Reynolds) seem to be mainly due to numerical errors, since they are much more acute if greater time steps are used for integration or smaller ensembles are considered (data not shown).
We also observe, and this is a limitation of the present model, that statistics are not fully consistent with those expected from isotropic turbulence. Indeed, for a homogeneous and isotropic velocity field, we expect that $\E[\varphi]=2\E[\tr(\tS\tS^\top)]=2\E[\tr(\tO\tO^\top)]$ \citep{Pop00}, a relation which is not perfectly met by the model. Figure \ref{fig:StatMeasures}(c) shows that the average enstrophy is a bit too small, with $2\E[\tr(\tO\tO^\top)]\approx 0.88 \E[\varphi]$, while the average dissipation a bit too high, $2\E[\tr(\tS\tS^\top)]\approx 1.17 \E[\varphi]$, compared to the average pseudo-dissipation. We discuss possible improvements of the model in order to correct this bias in section \ref{Sec:Conclu}.

Next, we study in figure \ref{fig:StatMeasures}(d) the correlation structure of the fields $\varphi(t)$ and $\varphi_{\tau_\eta}(t)$. With this aim, we estimate the autocovariance
\begin{equation}\label{eq:DefEcMeas}
\E_{c}\left[\log \varphi(t)\log \varphi(t+\tau)\right] = \E\left[\left(\log \varphi(t)-\E[\log \varphi]\right)\left(\log \varphi(t+\tau)-\E[\log \varphi]\right)\right],
\end{equation}
which, by stationarity, should be time independent. We know for sure, in the asymptotic regime $\tau_\eta\to 0$, that when it comes to the field $\varphi_{\tau_\eta}$ this correlation function behaves logarithmically with the time lag $\tau$, i.e.,
\begin{equation}\label{eq:BehavEcPhiTauEta}
\lim_{\tau_\eta/T\to 0}\E_{c}\left[\log \varphi_{\tau_\eta}(t)\log \varphi_{\tau_\eta}(t+\tau)\right] \build{\sim}_{\tau/T\to 0}^{} \mu^l \log\left(\frac{T}{\tau}\right).
\end{equation}
We compare in figure \ref{fig:StatMeasures}(d) the estimation of the autocovariance $\E_c$ \refp{eq:DefEcMeas} for the field $\varphi_{\tau_\eta}$ with the asymptotic logarithmic behaviour \refp{eq:BehavEcPhiTauEta}. It is seen that, as $\tau_\eta/T$ gets smaller, the autocovariance becomes more consistent with the expected logarithmic behaviour. As for $\varphi$, it is found that its corresponding centred correlation function follow very closely those of $\varphi_{\tau_\eta}$, for all the ratios $\tau_\eta/T$ here considered. We may nonetheless remark that, once again, improved time resolutions and larger ensembles were decisive in establishing this numerical trend.

\subsection{Joint PDF of the two invariants $R$ and $Q$}

\begin{figure}
\begin{center}
\input{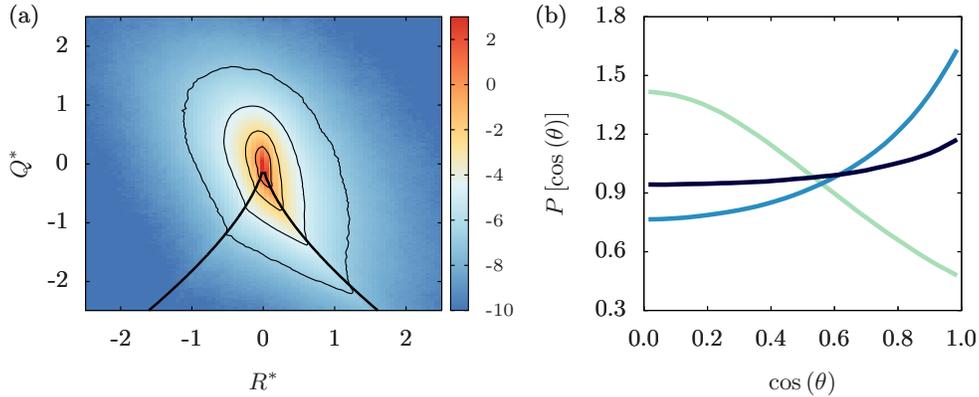}
\end{center}
\caption{\label{fig:RQsAlign} (a) Logarithmic representation of the joint PDF $\mathcal P(Q^*,R^*)$ of $R^* = R/\langle S_{ij}S_{ij}\rangle^{3/2}$ and $Q^*= Q/\langle S_{ij}S_{ij}\rangle$ computed from the statistics of $\tA$ at $\log(\tau_\eta/T)=-5.67$. Contour lines logarithmically spaced by factors of 1/10, starting at 1 near the origin. The thick line represents the zero-discriminant (or Vieillefosse) line: $(27/4)R^2+Q^3=0$. (b) PDF of the cosine of the angle between vorticity and the eigendirections of strain associated to the most compressive, intermediate and most extensive eigenvalues (lighter to darker), for the same $\tau_\eta/T$. }
\end{figure}

As an important characterization of the velocity gradient tensor $A_{ij}$, we study its two non vanishing invariants. The second invariant $Q$ is given by
\begin{equation}\label{eq:DefQ}
Q = -\frac{1}{2}\tr(\tA^2) = \frac{1}{4} |\bs{\omega}|^2-\frac{1}{2} \tr(\tS^2)
\end{equation}
where $\bs{\omega}$ is the vorticity vector and $\tS$ the symmetric part of $\tA$, and can be interpreted as the competition between enstrophy and dissipation (per unit viscosity). Therefore, positive $Q$ represents rotation-dominated
regions and negative $Q$ dissipation-dominated regions. Analogously, the third invariant $R$ is given by
\begin{equation}\label{eq:DefR}
R= -\frac{1}{3}\tr(\tA^3) = -\frac{1}{4} \omega_iS_{ij}\omega_j-\frac{1}{3} \tr(\tS^3),
\end{equation}
representing competition between \textit{enstrophy production} and \textit{dissipation production} \citep[we refer to][for discussions on this topic]{Tsi01,Wal09,Men11}. A numerical estimation of the PDF of the invariants $Q$ and $R$ is represented in figure \ref{fig:RQsAlign}(a). One may see that it is more elongated along the right tail of the Vieillefosse line and in the upper-left quadrant, precisely as observed in direct numerical simulations.

\subsection{Alignments of vorticity with the eigenframe of the deformation rate}
Another striking property of turbulence is the preferential alignment of vorticity with the strain eigendirection associated to the intermediate eigenvalue. We refer again to \cite{Tsi01,Wal09,Men11} for further discussions. Figure \ref{fig:RQsAlign}(b) displays the probability density function of the cosine of the angle $\theta$ between vorticity and the strain eigenvectors. It indicates the preferential alignment of vorticity with the intermediate eigendirection of deformation, as it is observed in DNS. We can add that the alignments PDFs (figure \ref{fig:RQsAlign}b) do not depend significantly on the $\tau_\eta/T$ value in the studied range, as well as the $RQ$ plane from figure \ref{fig:RQsAlign}(a), where the contour lines just get a bit more spaced as $\mathcal R_e$ increases due to additional non-Gaussianity (data not shown).

\subsection{Higher moments of the gradients and multifractal properties}\label{eq:NumStudyHO}
\begin{figure}
\begin{center}
\input{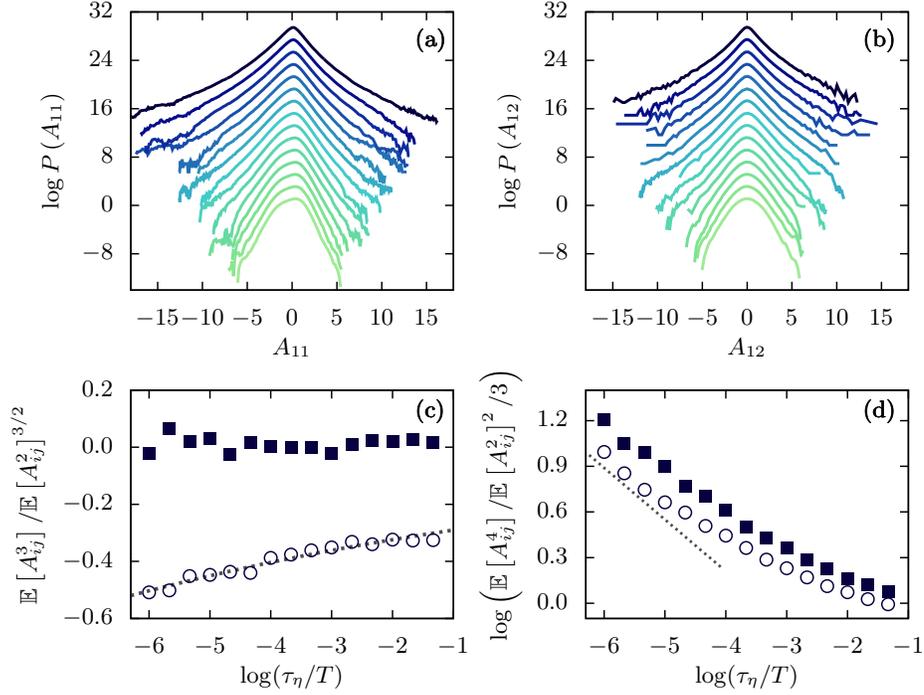}
\end{center}
\caption{\label{fig:Intermitt} (a) PDFs of diagonal elements $A_{11}$ in logarithm scale renormalized to unity variance and arbitrarily vertically shifted for clarity. Spanned $\log(\tau_\eta/T)$ values range from -1.33 to -5.67 with increments of -1/3 (from lighter to darker). (b) Similar study for the off-diagonal elements $A_{12}$. (c) Skewness and (d) flatness of both diagonal ($\modcirc$) and off-diagonal ($\blacksquare$) components. Predicted power laws \refp{eq:Skew} and \refp{eq:FlatExpIsh} are shown (dotted curves) for comparison.}
\end{figure}

We now characterize and quantify the intermittent properties of the statistics of $\tA$ in a sense that will become more precise.

First we estimate the PDF of the diagonal and off-diagonal elements of $\tA$. They are portrayed respectively in figures \ref{fig:Intermitt}(a) and \ref{fig:Intermitt}(b). All PDFs are normalized to unit variance and are arbitrarily vertically shifted for the sake of clarity. It is clear that in both cases, as the ratio $\tau_\eta$ decreases, the PDF undergoes a continuous shape deformation, from a Gaussian shape at large ratios towards large tails at small ratios. It is additionally noticeable that the cores of the diagonal elements PDFs (figure \ref{fig:Intermitt}(a)) are not symmetrical. This continuous shape deformation of the PDFs as the Reynolds number increases is a hallmark of multifractal behaviour.

To further characterize the intermittency phenomenon and the non-Gaussian behaviour of fluctuations, we compute the skewness $\E [A_{ij}^3]/(\E [A_{ij}^2])^{3/2}$ of diagonal and off-diagonal elements. It is seen from figure \ref{fig:Intermitt}(c) that, up to statistical uncertainties, only diagonal elements are skewed, as allowed by isotropy. Moreover, we contemplate that the skewness of diagonal elements exhibits a Reynolds number dependence (through the ratio $\tau_\eta/T$), as expected from multifractal phenomenologies. We superimpose in figure \ref{fig:Intermitt}(c) the power law (dotted line) observed in experimental and numerical data \citep[see equation 4.4 in][]{IshKan07}
\begin{align}\label{eq:Skew}
\frac{\E [ A^3_{11}]}{(\E[A^2_{11} ])^{3/2}}   \propto -\left(\frac{T}{\tau_\eta}\right)^{0.11},
\end{align}
which is very close to the prediction of the multifractal formalism \citep[see for instance][]{Fri95,CheCas12}. It is worth noting that in figure \ref{fig:Intermitt}(c) the power law \refp{eq:Skew} has been fitted beforehand on a log-log plot in order to fix the suitable multiplicative constant (not shown).

In the same spirit, we pursue the quantification of the level of intermittency through an estimate of the velocity gradients flatness, i.e. $\E [ A^4_{ij}]/(\E[A^2_{ij} ])^{2}$, again for both diagonal and off-diagonal elements of $\tA$. The resulting  behaviours are shown in figure \ref{fig:Intermitt}(d). It is evident that the depicted flatnesses exceed the Gaussian value $3$. We superimpose in this representation (dotted lines) the power law observed in numerical and experimental flows \citep{IshKan07}
\begin{align}\label{eq:FlatExpIsh}
\frac{\E [ A^4_{11}]}{(\E[A^2_{11} ])^{2}}   \propto \left(\frac{T}{\tau_\eta}\right)^{0.34},
\end{align}
which is also very close to what is predicted by the multifractal formalism \citep{Fri95,CheCas12}. For our simulated process \refp{eq:DynAFinal}, the power law \refp{eq:FlatExpIsh} seems to be representative of the statistics only for the smallest values of $\tau_{\eta}/T$ tested.

We emphasize that these numerically observed results for the velocity gradients higher-order moments illustrate the self-consistency of the present approach, once its single free parameter $\mu^l$ is given. Indeed, the power-law exponents as given in \refp{eq:Skew} and \refp{eq:FlatExpIsh} can be readily related to the intermittency coefficient $\mu^l$ of the Lagrangian framework since it governs the statistics of the overall amplitude of the gradients, and more precisely pseudo-dissipation.  They can also be derived, in the Eulerian framework, from the respective intermittency coefficient $\mu^e$ (see \refp{eq:MCPDEul}) as it is traditionally done in the literature \citep{Fri95}. We see that arguments developed by \cite{Bor93}, as recalled in section \ref{sec:ParamNum}, which relate both intermittency coefficients as $3\mu^e=2\mu^l$, give a consistent picture of the power-laws announced in \refp{eq:Skew} and \refp{eq:FlatExpIsh}.

\subsection{Rotation rate of anisotropic particles}\label{Sec:RRAP}

\begin{figure}
\begin{center}
\input{fig6.tex}
\end{center}
\caption{\label{fig:PDots} Variance of the rotation rate of anisotropic particles as a function of the aspect ratio $\alpha$ (see equation \ref{eq:JefEqu}). Solid line: DNS \citep{LiPer08}. Dotted line: RFD approximation \citep{CheMen13}. Symbols: predictions from the present model \refp{eq:DynAFinal}. Displayed $\tau_\eta/T$ ratios correspond to $\log(\tau_\eta/T) = -3.67 (\modcirc), -4.67 (\triangle)$ and $-5.67 (\square)$.}
\end{figure}

Many numerical and theoretical studies employ the Jeffery equation \citep{Jef22} to predict the time evolution of the orientation of axisymmetric ellipsoidal particles as they are advected and influenced by a turbulent velocity field. Specifically, Jeffery's equation for the unit orientation vector $\bp(t)$ in the ellipsoid major axis reads
\begin{align}\label{eq:JefEqu}
\frac{dp_n}{dt}=\Omega_{nj}p_j+ \lambda (S_{nj}p_j-p_np_kS_{kl}p_l),
\end{align}
where $\tS$ and $\tO$ are the strain and rotation rate tensors respectively, and $\lambda=(\alpha^2-1)/(\alpha^2+1)$, being $\alpha$ the particle's aspect ratio. In this convention, the limit $\alpha\to \infty$ corresponds to rods (elongated particles along the axis of symmetry), whereas discs are given in the limit $\alpha\to 0$. See for instance \cite{CheMen13} or \cite{GusEin14} for details, and \cite{VotSol17} for a recent review on the subject. It was proposed by \cite{ParCal12} to quantify, in experiments and numerical simulations, the variance of the rotation rate of anisotropic particles $\E[|\dot{\bp}|^2]$ as a function of the aspect ratio $\alpha$, which turns out to be a precise and demanding way to test models for the velocity gradient tensor $\tA$ \citep{CheMen13}.

We reproduce in figure \ref{fig:PDots} the behaviour of the normalized variance of the rotation rate $\E[|\dot{\bp}|^2]/\E[\varphi]$ as estimated in DNS \citep{LiPer08} and in trajectories built from the RFD approximation \citep[see][for details]{CheMen13}, along with the outcome of the present stochastic model \refp{eq:DynAFinal} for different $\tau_\eta/T$. Let us recall some of the interpretations that can be drawn from the behaviour of the rotation rate as a function of the aspect ratio $\alpha$. At high aspect ratios $\alpha\gg 1$, corresponding to elongated particles along the axis of symmetry (or \textit{rods}), the dynamics of anisotropic particles resembles that of material lines \citep[see for instance][for a review on this matter]{GuaLut05}. Therefore, a strong correlation between vorticity and the orientation vector is expected, leading in particular to a preferential alignment between them, as observed in numerical simulations \citep{PumWil11}. At low aspect ratios $\alpha\ll 1$, corresponding to disc shaped particles in the plane perpendicular to the orientation vector, it was observed that, contrary to rods, the orientation vector gets preferentially perpendicular to vorticity \citep{CheMen13,GusEin14}. As a consequence of this preferential alignment it is expected that discs will tumble, implying a higher rotation rate than in the rod case. This is what is observed in DNS \citep{ParCal12}: $\E[|\dot{\bp}|^2]$ increases monotonically as the aspect ratio $\alpha$ decreases from rods ($\alpha\gg 1$) to discs   ($\alpha\ll 1$).

It was shown in \cite{CheMen13} that the stochastic model for velocity gradients coined with the RFD approximation is only able to satisfactorily reproduce the behaviour of the rotation rate observed in DNS for aspect ratios larger than unity. To this regard, the present model \refp{eq:DynAFinal} exhibits an even more striking agreement for these aspect ratios. For particles with smaller aspect ratios ($\alpha <1$), it is seen that the RFD approximation predicts a decrease of the rotation rate, whereas in DNS it keeps increasing up to a saturation value. The present model \refp{eq:DynAFinal} shows a similar saturation structure of the fluctuations with the aspect ratio $\alpha$, an improvement in the modelling of the rotation rate of anisotropic particles compared to the RFD approximation, but overall, even if the present model \refp{eq:DynAFinal} exhibits many aspects of the statistical properties of the velocity gradient tensor $\tA$, it still fails at providing a satisfying picture of the tumbling motions of discs in turbulence. Let us finally remark that the predictions of the present model on the rotation rate are almost independent of the Reynolds number (or equivalently $\tau_\eta/T$), whereas it is shown in former sections that the statistics of $\tA$ do have a realistic dependence. This is consistent with the trends observed from DNS studies \citep{ParCal12}.

%%%%%%%%%%%%%%%%%%
%%  NEW SECTION %%
%%%%%%%%%%%%%%%%%%

\section{Conclusion and perspectives}\label{Sec:Conclu}

We have proposed a new stochastic time evolution of the velocity gradient tensor $\tA$ along a Lagrangian trajectory \refp{eq:DynAFinal}, and studied numerically the statistical properties of its solution.

The main ingredients of this model are (i) a regularization by the joint action of closures for pressure Hessian and the viscous Laplacian of the exact self-stretching term entering the dynamics \refp{eq:RFDDeterm} and (ii) an additional stochastic closure that constrains the pseudo-dissipation to follow as close as possible the dynamics of a multifractal measure. Doing so, we end up with a non-Markovian dynamics that depends on the Reynolds number (through the ratio $\tau_\eta/T$) and on a free parameter $\mu^l$, the intermittency coefficient, taken to be equal to $0.3$ as observed in DNS of homogeneous and isotropic turbulence. To our knowledge, this is the first stochastic proposition for the dynamics of A which incorporates the intermittency parameter $\mu^l$ and is able to reproduce, in a realistic way, the statistics of the small scales of turbulence at any Reynolds numbers without relying on hierarchical structures for the couplings of multiscale velocity gradient tensors, as it was previously addressed by \cite{BifChe07} and, more recently, by \cite{JohMen17}.

To illustrate the realism of our model, we have numerically studied the statistical properties of its solution and showed that the variance and higher-order moments of the elements of $\tA$ behave accurately with the Reynolds number as observed in experimental and numerical data, and as captured by the multifractal phenomenology of turbulence. Furthermore, geometrical statistical properties such as vorticity alignments with respect to the eigenframe of deformation and joint distribution of the invariants are also reproduced in a satisfactory manner.

Despite the realism of the so-obtained statistical behaviour of $\tA$, and the theoretical success to gather in a certain sense the multifractal phenomenology and the velocity gradient dynamics, it is also shown that the tumbling phenomenon of anisotropic particles of small aspect ratios (i.e. discs) is still not accurately reproduced.

The present proposition \refp{eq:DynAFinal} could be improved in several ways. First of all, the statistical properties of $\tA$ from the model are not fully consistent with those of isotropic turbulence (see the behaviour of the averaged dissipation, enstrophy and pseudo-dissipation in figure \ref{fig:StatMeasures}(c), and the corresponding discussion provided in section \ref{sec:StatVarPhi}. To this regard, a first step toward an improvement of the model would be to explore the dependence of the closure made in \refp{eq:Diffabc} where the choice $c=0$ is made. Another possible way to improve these statistics would be to consider more sophisticated closures of pressure Hessian and viscous term other than the RFD approximation, such as the propositions of \cite{WilMen14} and, more generally, those of \cite{JohMen16}. In a different line of development, it would be interesting to try to adjust the model to more complex anisotropic flows, in the spirit of \cite{GirSpe95} for instance.

As for the statistics of rotation rate of anisotropic particles (section \ref{Sec:RRAP}), an alternative way to explore the implied physics of the tumbling phenomenon and the implications on velocity gradient statistics would be to quantify the gradients' third order correlation functions as suggested in \cite{GusEin14}. Such a study, supplemented by the analysis of homogeneous and isotropic turbulence DNS, would help us to better understand the underlying mechanisms leading to the tumbling phenomenon and to  design accurate stochastic models able to reproduce them. We keep these perspectives for future investigations.

\section*{Acknowledgements}
The authors thank the PSMN (P\^ole Scientifique de Mod\'elisation Num\'erique) computing centre of ENS de Lyon for numerical resources. R.M.P. acknowledges CAPES for financial support. L.M. thanks CNPq for partial support and the ENS-Lyon for its warm hospitality and great scientific environment during his collaboration visit. L.C. is supported by ANR grants \textsc{Liouville} ANR-15-CE40-0013 and \textsc{MultiFracs} ANR-16-CE33-0020.

%%%%%%%%%%%%%%%%%%
%%  NEW SECTION %%
%%%%%%%%%%%%%%%%%%

\appendix

\section{Exponentiation of an Ornstein-Uhlenbeck process: definition and consequences}\label{ann:ExpOU}

\subsection{Definition and moments of the pseudo-dissipation field}

\cite{Pop90} has proposed to model the instance of the pseudo-dissipation field as seen by a Lagrangian particle as
\begin{equation}\label{eq:SDEGPDiss}
d\varphi(t)= \varphi(t)\left[\hat{a}^2-\frac{1}{T}\log[\tau_{\eta}^2\varphi(t)] \right]dt+2\hat{a}\varphi(t) W(dt),
\end{equation}
where $\hat{a}$ is the free parameter of the model. The dynamics defined by \refp{eq:SDEGPDiss} has a unique statistically stationary solution. Without loss of generality, one may choose some arbitrary positive initial condition and take a shortcut to the stationary regime considering formally $t_0=-\infty$. This unique solution reads
\begin{equation}\label{eq:ExpOUDiss}
\varphi(t) = \frac{1}{\tau_{\eta}^2}\exp\left[-\hat{a}^2T+2\hat{a}\int_{-\infty}^t e^{-\frac{t-s}{T}}W(ds)\right].
\end{equation}
It is easily seen from \refp{eq:ExpOUDiss} that moments of $\varphi$ are time-independent (stationary regime) and, considering for simplicity $q\in\mathbb N$, given by
\begin{equation}\label{eq:MomqExpOUDiss}
\E(\varphi^q)=\frac{1}{\tau_{\eta}^{2q}}e^{\hat{a}^2Tq(q-1)}.
\end{equation}
In particular, Kolmogorov's prediction \refp{eq:KolPredPD} is automatically fulfilled in this formalism, that is, $\E(\varphi)=1/\tau_{\eta}^{2}$.

Let us now compute the moments of the coarse-grained pseudo-dissipation $\varphi_\tau$ \refp{eq:MeanTauDiss} along a Lagrangian trajectory. We have, for $q\in\mathbb N^*$,
$$ \E{\varphi_\tau^q}=\frac{1}{\tau^q}\int_{[t-\tau,t]^q}\E\left[\prod_{n=1}^q \varphi(s_n)\right]\prod_{n=1}^qds_n.$$
Notice that
$$ \tau_{\eta}^{2q}\prod_{n=1}^q \varphi(s_n) = e^{-q\hat{a}^2T}\exp\left[2\hat{a}\sum_{n=1}^{q}\int_{-\infty}^{s_n} e^{-\frac{s_n-s}{T}}W(ds)\right],$$
where each of the stochastic integrals in the argument of the exponential is a zero-average Gaussian process of variance
$$ \E\left[\left(\int_{-\infty}^{s_n} e^{-\frac{s_n-s}{T}}W(ds)\right)^2\right]=\frac{T}{2}$$
and covariance
$$ \E\left[\left(\int_{-\infty}^{s_i} e^{-\frac{s_i-s}{T}}W(ds)\right)\left(\int_{-\infty}^{s_j} e^{-\frac{s_j-s}{T}}W(ds)\right)\right]=\frac{T}{2}e^{-\frac{|s_i-s_j|}{T}}.$$
Using now that $\E(e^g)=e^{\frac{1}{2}\E(g^2)}$ for any zero-average Gaussian variable $g$, we have
\begin{equation}\label{eq:VarPhiTauqOU}
 \E{\varphi_\tau^q}=\frac{1}{\tau_{\eta}^{2q}}\frac{1}{\tau^q}\int_{[0,\tau]^q}\exp\left[2\hat{a}^2T\sum_{i<j=1}^q e^{-\frac{|s_i-s_j|}{T}}\right]\prod_{n=1}^qds_n.
\end{equation}
Rescaling the dummy variables $s_i$ by $\tau$ and taking the limit $\tau\to 0$ then recovers the $q^{\mbox{th}}$-order moment of $\varphi$ expressed in \refp{eq:MomqExpOUDiss}.

\subsection{Discussions on the Reynolds number dependence of the free parameter of the theory}

To be consistent with both observed fluctuations of turbulence and the multifractal formalism \citep{Bor93,Fri95}, we expect at the very least a power law dependence of the $\varphi$ moments \refp{eq:ExpOUDiss} with the Reynolds number $\mathcal R_e$. In this spirit, \cite{GirPop90} suggest to set the free parameter $\hat{a}$, up to a Reynolds independent additive constant, equal to
\begin{equation}\label{eq:Choiceabar}
\hat{a}^2=\frac{\mu^l}{2T}\log\left( \frac{T}{\tau_\eta}\right),
\end{equation}
defining thus a dimensionless free parameter $\mu^l$ known in the language of the multifractal formalism as the intermittency coefficient.  Hence, recalling that the ratio $T/\tau_\eta$ is proportional to $\sqrt{\mathcal R_e}$, we see that modelling a pseudo-dissipation trajectory as the exponential of an Ornstein-Uhlenbeck process \refp{eq:SDEGPDiss} predicts in a consistent way the Reynolds number dependence of the higher-order moments \refp{eq:MomqExpOUDiss} if the free parameter is chosen with an appropriate Reynolds number dependence \refp{eq:Choiceabar}.

Let us now discuss the applicability of the Refined Similarity Hypothesis (RSH) in this context. Recall that RSH bridges in a Eulerian context the statistics of the coarse-grained dissipation with the statistics of the spatial velocity increments \citep{Fri95}. In a Lagrangian context \citep{Bor93}, but using pseudo-dissipation instead of dissipation, the RSH implies in a similar way \citep[see][for details]{CheCas12} that the statistics of temporal velocity increments is given by that of an appropriate power of $\varphi_{\tau}$. Among other features, if the statistics of the increments is Reynolds number independent in the inertial range, as observed in real flows, this also holds for $\tau_\eta^2\varphi_{\tau}$. It is then easy to see, since the parameter $\hat{a}$ diverges with the Reynolds number \refp{eq:Choiceabar}, that the high-order moments of $\tau_\eta^2\varphi_{\tau}$ \refp{eq:VarPhiTauqOU} will also diverge for any $q>1$. This is not consistent with experimental observations showing that velocity fluctuations are independent of the viscosity in the inertial range. Thus, to this regard, the choice of an exponentiated Ornstein-Uhlenbeck process \refp{eq:SDEGPDiss} as a model for pseudo-dissipation trajectories is not consistent with the standard phenomenology. We will see in the following that the multifractal model we consider indeed predicts simultaneously an appropriate Reynolds number dependence for the $\varphi$ moments \refp{eq:MomqExpOUDiss} and an appropriate behaviour of the moments of $\tau_\eta^2\varphi_{\tau}$ \refp{eq:VarPhiTauqOU} as the Reynolds number tends to infinity.

%%%%%%%%%%%%%%%%%%
%%  NEW SECTION %%
%%%%%%%%%%%%%%%%%%

\section{A causal multiplicative chaos: Definition and statistical properties}
\label{Ann:MultiChaosDiss}
\subsection{Setup and notations}

We consider the following linear stochastic differential equation
\begin{equation}\label{eq:sdeX}
dX_{\tau_\eta}(t) = \left[-\frac{1}{T} X_{\tau_\eta}(t)+\beta_{\tau_\eta}(t)\right]dt + \frac{1}{\sqrt{{\tau_\eta}}}W(dt),
\end{equation}
with $W$ a Gaussian white noise and $\beta_{\tau_\eta}(t)$ a random function known in the literature as a (${\tau_\eta}$-regularized) fractional Gaussian noise of Hurst exponent $H=0$ \citep{ManVan68}:
\begin{equation}\label{eq:beta}
\beta_{\tau_\eta}(t) = -\frac{1}{2} \int_{s=-\infty}^t \frac{1}{(t-s+{\tau_\eta})^{3/2}} W(ds).
\end{equation}
Equation \ref{eq:sdeX} unique solution may be conveniently written as
\begin{equation}\label{eq:X}
X_{\tau_\eta}(t) = \int_{s=-\infty}^t e^{-\frac{t-s}{T}}\beta_{\tau_\eta}(s)ds + \frac{1}{\sqrt{{\tau_\eta}}}\int_{s=-\infty}^t e^{-\frac{t-s}{T}}W(ds),
\end{equation}
if we choose, formally, the initial condition $X_{\tau_\eta}(-\infty)=0$, taking a shortcut to define these Gaussian processes (\ref{eq:X} and \ref{eq:beta}) starting from an infinitely ancient time. As we will see, this makes perfect sense and leads to finite variance processes (at a finite $\tau_\eta$). This enables us to work directly with a causal and stationary framework. From a numerical point of view, it is equivalent to start from any initial condition $X(t_0)<+\infty$ or, similarly, truncate the stochastic integrals from \refp{eq:X} and \refp{eq:beta} over $[t_0,t]$ and propagate in time until reaching a stationary regime (see also the discussion provided in section \ref{sec:Num}). For convenience, we define the sub processes $X_{\tau_\eta}^{(1)}(t)$ and $X_{\tau_\eta}^{(2)}(t)$, such that $X_{\tau_\eta}(t) = X_{\tau_\eta}^{(1)}(t)+X_{\tau_\eta}^{(2)}(t)$, as
\begin{equation}\label{eq:X1}
X_{\tau_\eta}^{(1)}(t) = \int_{s=-\infty}^t e^{-\frac{t-s}{T}}\beta_{\tau_\eta}(s)ds
\end{equation}
which is known in the literature as a fractional Ornstein-Uhlenbeck (FOU) process \citep{CheKaw03} of Hurst exponent $H=0$, and
\begin{equation}\label{eq:X2}
X_{\tau_\eta}^{(2)}(t) =\frac{1}{\sqrt{{\tau_\eta}}}\int_{s=-\infty}^t e^{-\frac{t-s}{T}}W(ds)
\end{equation}
a standard Ornstein-Uhlenbeck (OU) process.

A process similar to $X_{\tau_\eta}$, although different, has been considered heuristically by \cite{Sch03}. In our formulation \refp{eq:sdeX}, the integral time scale $T$ comes into the dynamics through the damping term, which allows us to derive rigorously, in the stationary regime, its statistical properties. We preserve from the heuristics of \cite{Sch03} the correlated OU-process $X_{\tau_\eta}^{(2)}$ \refp{eq:X2} which plays a crucial role in the determination of the statistics in the stationary regime.

\subsection{Variance and covariance in the stationary regime}
The Gaussian process $X_{\tau_\eta}(t)$ \refp{eq:X} reaches a stationary regime, with
\begin{equation}\label{eq:ResVar}
\E \left[\left(X_{\tau_\eta}\right)^2\right] =\int_{\mathbb R^+}e^{-\frac{h}{T}}g_{\tau_\eta}(h)dh,
\end{equation}
and
$$g_{\tau_\eta}(h)=\frac{1}{h+\tau_\eta+\sqrt{\tau_\eta(h+\tau_\eta)}}. $$
Furthermore, in the asymptotic limit of infinite Reynolds number, i.e. ${\tau_\eta} \to 0$, we have the following logarithmic diverging behaviour
\begin{align}\label{eq:TheoAsympVarX}
\E \left[\left(X_{\tau_\eta}\right)^2\right] \build{\sim}_{{\tau_\eta} \to 0}^{}\log\left(\frac{T}{{\tau_\eta}}\right).
\end{align}
In the asymptotic limit of infinite Reynolds number $\tau_\eta\to 0$, as opposed to the variance, the covariance remains bounded for $\tau>0$, and we note
\begin{equation}\label{eq:ResCovarAsympt}
\E \left[X(t)X(t+\tau)\right]=\lim_{\tau_\eta\to 0}\E \left[X_{\tau_\eta}(t)X_{\tau_\eta}(t+\tau)\right].
\end{equation}
In the limit of vanishing time lag $\tau$, we have moreover the following logarithmic divergence of the correlation function
\begin{equation}\label{eq:ResCovarAsymptTau}
\E \left[X(t)X(t+\tau)\right]\build{\sim}_{\tau\to 0}^{} \log\left( \frac{T}{\tau}\right).
\end{equation}

\subsection{Proofs}
\subsubsection{The Ornstein-Uhlenbeck process $X_{\tau_\eta}^{(2)}$}

It is well known that
$$ \E \left[X_{\tau_\eta}^{(2)}(t)\right]=0,$$
as readily seen from \refp{eq:X2}. As for the covariance, we arrive at
\begin{align}\label{eq:covarX2}
\E \left[X_{\tau_\eta}^{(2)}(t)X_{\tau_\eta}^{(2)}(t+\tau)\right] = \frac{1}{2}\frac{T}{\tau_\eta}e^{-\frac{|\tau|}{T}},
\end{align}
independently of $t$. Notice that by taking $\tau=0$ in \refp{eq:covarX2} the variance itself may be written as
\begin{align}\label{eq:varX2}
\E \left[\left(X_{\tau_\eta}^{(2)}(t)\right)^2\right] = \frac{1}{2}\frac{T}{\tau_\eta}.
\end{align}
We can see that both the covariance and the variance of $X_{\tau_\eta}^{(2)}$ diverge as the Reynolds number becomes infinite.

\subsubsection{The  fractional Ornstein-Uhlenbeck process $X_{\tau_\eta}^{(1)}$}

Similarly, it is straightforward to verify that
$$ \E \left[X_{\tau_\eta}^{(1)}(t)\right]=0.$$
The process' covariance is given by
\begin{equation}
\E \left[X_{\tau_\eta}^{(1)}(t)X_{\tau_\eta}^{(1)}(t+\tau)\right]=e^{-\frac{|\tau|}{T}}\int_{s_1=-\infty}^{t}\int_{s_2=-\infty}^{t+|\tau|}e^{-\frac{2t-s_1-s_2}{T}} C_{\tau_\eta}(s_1-s_2)ds_1ds_2,
\end{equation}
where $ C_{\tau_\eta}(h)$ stands for the covariance of the fractional noise $\beta_\tau$ \refp{eq:beta} and reads
\begin{equation}\label{eq:Ctaueta}
C_{\tau_\eta}(h) = \E\left[ \beta_{\tau_\eta}(0)\beta_{\tau_\eta}(h)\right] = \frac{1}{4}\int_{u=0}^{\infty}\frac{1}{(u+{\tau_\eta})^{3/2}}\frac{1}{(u+|h|+{\tau_\eta})^{3/2}}du.
\end{equation}
Therefore, it may be conveniently written as
\begin{align}\label{eq:covarX1}
\E \left[X_{\tau_\eta}^{(1)}(t)X_{\tau_\eta}^{(1)}(t+\tau)\right]&=e^{-\frac{|\tau|}{T}}\E \left[\left(X_{\tau_\eta}^{(1)}\right)^2\right]\notag \\
&+e^{-\frac{|\tau|}{T}}\int_{s_1=-\infty}^{t}\int_{s_2=t}^{t+|\tau|}e^{-\frac{2t-s_1-s_2}{T}} C_{\tau_\eta}(s_1-s_2)ds_1ds_2.
\end{align}
As for the variance, with the change of variables $t-s_i\rightarrow s_i$ for $i\in\{1,2\}$, we arrive at
$$\E \left[\left(X_{\tau_\eta}^{(1)}\right)^2\right]=\int_{(\mathbb R^+)^2}e^{-\frac{1}{T}(s_1+s_2)}C_{\tau_\eta}(s_1-s_2)ds_1ds_2,$$
that can be worked out a bit to become
\begin{align*}
\E \left[\left(X_{\tau_\eta}^{(1)}\right)^2\right]&=T\int_0^{\infty}e^{-\frac{h}{T}}C_{\tau_\eta}(h)dh\\
&= \frac{T}{4}\int_{(\mathbb R^+)^2}e^{-\frac{h}{T}}\frac{1}{(u+{\tau_\eta})^{3/2}}\frac{1}{(u+h+{\tau_\eta})^{3/2}}dudh.
\end{align*}
Focusing on the integration of the dummy variable $h$, an integration by parts leads to
\begin{align}\label{eq:varX1}
\E \left[\left(X_{\tau_\eta}^{(1)}\right)^2\right]&=\frac{T}{4}\left[ 2\int_0^{\infty}\frac{1}{(u+{\tau_\eta})^2}du -\frac{2}{T}\int_{(\mathbb R^+)^2}e^{-\frac{h}{T}}\frac{1}{(u+{\tau_\eta})^{3/2}}\frac{1}{(u+h+{\tau_\eta})^{1/2}}dudh \right]\notag \\
&= \frac{1}{2}\frac{T}{\tau_\eta} -\frac{1}{2}\int_{(\mathbb R^+)^2}e^{-\frac{h}{T}}\frac{1}{(u+{\tau_\eta})^{3/2}}\frac{1}{(u+h+{\tau_\eta})^{1/2}}dudh.
\end{align}

\subsubsection{Covariance of $X_{\tau_\eta}^{(1)}$ and $X_{\tau_\eta}^{(2)}$}

Without loss of generality, consider for instance $\tau\ge 0$. One has
$$ \E \left[X_{\tau_\eta}^{(1)}(t)X_{\tau_\eta}^{(2)}(t+\tau)\right]=\frac{1}{\sqrt{{\tau_\eta}}}e^{-\frac{\tau}{T}}\int_{s_1=-\infty}^{t}\int_{s_2=-\infty}^{t+\tau}e^{-\frac{1}{T} (2t-s_1-s_2)}\E \left[ \beta_{\tau_\eta}(s_1)W(ds_2)\right]ds_1,$$
where it is simple to give meaning to the following rule of calculation
\begin{align*}
 \E \left[ \beta_{\tau_\eta}(s_1)W(ds_2))\right]&=-\frac{1}{2} \int_{u=-\infty}^{s_1} \frac{1}{(s_1-u+{\tau_\eta})^{3/2}} \E \left[W(du)W(ds_2)\right]\\
&=-\frac{1}{2} \frac{1}{(s_1-s_2+{\tau_\eta})^{3/2}}1_{s_1\ge s_2}ds_2.
\end{align*}
Here $1_{s_1\ge s_2}$ is the indicator function of the ensemble $\{(s_1,s_2)\in\mathbb R^2, s_1\ge s_2\}$. Manipulating a bit, one gets
\begin{equation}\label{eq:X1X2tau}
\E \left[X_{\tau_\eta}^{(1)}(t)X_{\tau_\eta}^{(2)}(t+\tau)\right]=e^{-\frac{\tau}{T}}\E \left[X_{\tau_\eta}^{(1)}(t)X_{\tau_\eta}^{(2)}(t)\right],
\end{equation}
where
\begin{align*}
 \E \left[X_{\tau_\eta}^{(1)}(t)X_{\tau_\eta}^{(2)}(t)\right]&=-\frac{1}{2\sqrt{{\tau_\eta}}}\int_{]-\infty,t]^2}e^{-\frac{1}{T} (2t-s_1-s_2)}\frac{1}{(s_1-s_2+{\tau_\eta})^{3/2}}1_{s_1\ge s_2}ds_1ds_2\\
&=-\frac{1}{2\sqrt{{\tau_\eta}}}\int_{(\mathbb R^+)^2}e^{-\frac{1}{T} (s_1+s_2)}\frac{1}{(s_1-s_2+{\tau_\eta})^{3/2}}1_{s_1\ge s_2}ds_1ds_2\\
&=-\frac{1}{2\sqrt{{\tau_\eta}}}\frac{T}{2}\int_0^{\infty}\frac{e^{-\frac{h}{T}}}{(h+{\tau_\eta})^{3/2}}dh.
\end{align*}
An integration by parts on the remaining integral finally leads to
\begin{align}\label{eq:CorrX1X2}
\E \left[X_{\tau_\eta}^{(1)}(t)X_{\tau_\eta}^{(2)}(t)\right] = -\frac{1}{2}\frac{T}{\tau_\eta}+ \frac{1}{2\sqrt{{\tau_\eta}}}\int_0^{\infty}\frac{e^{-\frac{1}{T} h}}{(h+{\tau_\eta})^{1/2}}dh.
\end{align}
Similarly, recalling that $\tau\ge 0$, one may write
\begin{align}\label{eq:X1tauX2}
\E &\left[X_{\tau_\eta}^{(1)}(t+\tau)X_{\tau_\eta}^{(2)}(t)\right]=\frac{1}{\sqrt{{\tau_\eta}}}e^{-\frac{\tau}{T}}\int_{s_1=-\infty}^{t+\tau}\int_{s_2=-\infty}^{t}e^{-\frac{1}{T} (2t-s_1-s_2)}\E \left[ \beta_{\tau_\eta}(s_1)W(ds_2)\right]ds_1\notag \\
&=e^{-\frac{\tau}{T}}\left( \E \left[X_{\tau_\eta}^{(1)}(t)X_{\tau_\eta}^{(2)}(t)\right]+\frac{1}{\sqrt{{\tau_\eta}}}\int_{s_1=t}^{t+\tau}\int_{s_2=-\infty}^{t}e^{-\frac{1}{T} (2t-s_1-s_2)}\E \left[ \beta_{\tau_\eta}(s_1)W(ds_2)\right]ds_1\right).
\end{align}

\subsubsection{Variance of $X_{\tau_\eta}$}

Noticing that
$$\E \left[\left(X_{\tau_\eta}\right)^2\right] = \E \left[\left(X_{\tau_\eta}^{(1)}\right)^2\right]+\E \left[\left(X_{\tau_\eta}^{(2)}\right)^2\right]+2 \E \left[X_{\tau_\eta}^{(1)}X_{\tau_\eta}^{(2)}\right], $$
one obtains, using \refp{eq:varX2}, \refp{eq:varX1} and \refp{eq:CorrX1X2},
\begin{equation}\label{eq:preVarXtau}
\E \left[\left(X_{\tau_\eta}\right)^2\right] = -\frac{1}{2}\int_{\mathbb R^+}e^{-\frac{h}{T}}f_{\tau_\eta}(h)dh+ \frac{1}{\sqrt{{\tau_\eta}}}\int_0^{\infty}\frac{e^{-\frac{h}{T} h}}{(h+{\tau_\eta})^{1/2}}dh,
\end{equation}
where
\begin{equation}\label{eq:fdef}
f_{\tau_\eta}(h)=\int_{\mathbb R^+}\frac{1}{(u+{\tau_\eta})^{3/2}}\frac{1}{(u+h+{\tau_\eta})^{1/2}}du.
\end{equation}
Integrating \refp{eq:fdef} by parts results in
$$f_{\tau_\eta}(h)= \frac{2}{\sqrt{{\tau_\eta}}}\frac{1}{(h+{\tau_\eta})^{1/2}} -2g_{\tau_\eta}(h),$$
with
\begin{equation}\label{eq:ExactIntPL}
g_{\tau_\eta}(h)=\frac{1}{2}\int_{\mathbb R^+}\frac{1}{(u+{\tau_\eta})^{1/2}}\frac{1}{(u+h+{\tau_\eta})^{3/2}}du=\frac{1}{h+\tau_\eta+\sqrt{\tau_\eta(h+\tau_\eta)}}.
\end{equation}
Now, inserting this into the variance \refp{eq:preVarXtau}, one finally derives
\begin{align}\label{eq:VarXtau}
\E \left[\left(X_{\tau_\eta}\right)^2\right] =\int_{\mathbb R^+}e^{-\frac{h}{T}}g_{\tau_\eta}(h)dh,
\end{align}
which entails the proposition made in \refp{eq:ResVar}.

\subsubsection{Asymptotics in the limit of infinite Reynolds numbers ${\tau_\eta}\to 0$}\label{Sec:AsyVarRe}

We rewrite the process' variance rescaling the integration variable by ${\tau_\eta}$, obtaining
$$\E \left[\left(X_{\tau_\eta}\right)^2\right] =\int_{\mathbb R^+}e^{-\frac{\tau_\eta}{T}  h}r(h)dh,$$
with
$$ r(h)=\frac{1}{h+1+\sqrt{h+1}}.$$
Notice that $r$ is a bounded function of its argument, in particular, $r(0)=1$ and, as $h\to \infty$, it behaves as
\begin{align}\label{eq:Asymptrh}
 r(h) = \frac{1}{h}-\frac{1}{h^{3/2}}+o\left( \frac{1}{h^{3/2}}\right).
\end{align}
To deduce the behaviour of the variance of the process $X_{\tau_\eta}$ \refp{eq:VarXtau} as $\tau_\eta\to 0$, we split the integral into two contributions, namely
$$\E \left[\left(X_{\tau_\eta}\right)^2\right] =\int_{0}^1e^{-\frac{\tau_\eta}{T} h}r(h)dh+\int_{1}^{\infty}e^{-\frac{\tau_\eta}{T} h}r(h)dh.$$
The first contribution is bounded with ${\tau_\eta}$ and tends to $\int_{0}^1 r(h)dh$ when ${\tau_\eta}\to 0$. Nevertheless, the second one will diverge when ${\tau_\eta}\to 0$ and will thus dominate. To see how fast it diverges, we write it as
$$\int_{1}^{\infty}e^{-\frac{\tau_\eta}{T}  h}r(h)dh =\int_{1}^{\infty}e^{-\frac{\tau_\eta}{T} h}\left[r(h) -\frac{1}{h}\right]dh + \int_{1}^{\infty}e^{-\frac{\tau_\eta}{T} h}\frac{1}{h}dh. $$
Using the expansion \refp{eq:Asymptrh} we see that $r(h)-\frac{1}{h}$ is integrable when $h\to \infty$, we can thus apply Lebesgue's dominated convergence theorem and conclude that the first contribution is bounded with ${\tau_\eta}$ (and tends to $\int_{1}^{\infty}\left[r(h) -\frac{1}{h}\right]dh$ when ${\tau_\eta} \to 0$). Only the second contribution will diverge when ${\tau_\eta} \to 0$, so we can write
$$ \E \left[\left(X_{\tau_\eta}\right)^2\right] \build{\sim}_{{\tau_\eta} \to 0}^{} \int_{1}^{\infty}e^{-\frac{1}{T} {\tau_\eta} h}\frac{1}{h}dh.$$
To assess how this well behaved quantity diverges, we integrate by parts to observe that
\begin{align*}
\int_{1}^{\infty}e^{-\frac{\tau_\eta}{T} h}\frac{1}{h}dh&=\frac{\tau_\eta}{T}\int_{1}^{\infty}e^{-\frac{\tau_\eta}{T}  h}\log(h)dh = \int_{\frac{\tau_\eta}{T}}^{\infty}e^{-u}\log\left(\frac{T}{\tau_\eta}u\right)dh\\
&\build{\sim}_{{\tau_\eta} \to 0}^{}\log\left(\frac{T}{{\tau_\eta}}\right),
\end{align*}
which concludes the proof of the proposition made in \refp{eq:TheoAsympVarX}.

\subsubsection{Asymptotics of the covariance of $X_{\tau_\eta}$}\label{Ann:AsymptCovXtaueta}
Starting from
\begin{align*}
\E \left[X_{\tau_\eta}(t)X_{\tau_\eta}(t+\tau)\right] &= \E \left[X_{\tau_\eta}^{(1)}(t)X_{\tau_\eta}^{(1)}(t+\tau)\right] +\E \left[X_{\tau_\eta}^{(2)}(t)X_{\tau_\eta}^{(2)}(t+\tau)\right] \\\
&+ \E \left[X_{\tau_\eta}^{(1)}(t)X_{\tau_\eta}^{(2)}(t+\tau)\right]+\E \left[X_{\tau_\eta}^{(1)}(t+\tau)X_{\tau_\eta}^{(2)}(t)\right],
\end{align*}
one has, using \refp{eq:covarX2}, \refp{eq:covarX1}, \refp{eq:X1X2tau} and \refp{eq:X1tauX2},
\begin{align}\label{eq:covarXtau1}
\E &\left[X_{\tau_\eta}(t)X_{\tau_\eta}(t+\tau)\right] = e^{-\frac{|\tau|}{T}}\E \left[(X_{\tau_\eta})^2\right] \nonumber\\
&+ e^{-\frac{|\tau|}{T}}\int_{s_1=0}^{\infty}\int_{s_2=-|\tau|}^{0}e^{-\frac{s_1+s_2}{T}}\left[ C_{\tau_\eta}(s_1-s_2)-\frac{1}{2\sqrt{\tau_\eta}}\frac{1}{(s_1-s_2+\tau_\eta)^{3/2}}\right]ds_1ds_2.
\end{align}
Let's denote the second term of the RHS of the former equation by $I_{\tau_\eta}(\tau)$. Notice first that one may integrate by parts the definition of $C_{\tau_\eta}$ \refp{eq:Ctaueta} to stablish
$$ C_{\tau_\eta}(h)=\frac{1}{2\sqrt{\tau_\eta}}\frac{1}{(|h|+\tau_\eta)^{3/2}}-\frac{3}{4}\int_0^{\infty}\frac{1}{(v+\tau_\eta)^{1/2}}\frac{1}{(v+|h|+\tau_\eta)^{5/2}}dv,$$
that may be used to obtain
\begin{align*}
I_{\tau_\eta}(\tau) &= e^{-\frac{|\tau|}{T}}\int_{s_1=0}^{\infty}\int_{s_2=-|\tau|}^{0}e^{-\frac{s_1+s_2}{T}}\left[ C_{\tau_\eta}(s_1-s_2)-\frac{1}{2\sqrt{\tau_\eta}}\frac{1}{(s_1-s_2+\tau_\eta)^{3/2}}\right]ds_1ds_2\\
&=-\frac{3}{4}e^{-\frac{|\tau|}{T}}\int_{s_1=0}^{\infty}\int_{s_2=-|\tau|}^{0}\int_{v=0}^{\infty}e^{-\frac{s_1+s_2}{T}}\frac{1}{(v+\tau_\eta)^{1/2}}\frac{1}{(v+s_1-s_2+\tau_\eta)^{5/2}}ds_1ds_2dv.
\end{align*}
With the change of variables $h=s_1-s_2$ (while keeping the integration over $s_1$ and $v$), we perform a integration by parts over $h$, manipulate a bit, and arrive at
\begin{align*}
I_{\tau_\eta}(\tau) &= -e^{-\frac{|\tau|}{T}}\E \left[(X_{\tau_\eta})^2\right] \\
&+ \frac{1}{2}\int_{s_1=0}^{\infty}\int_{v=0}^{\infty}e^{-\frac{s_1}{T}}\frac{1}{(v+\tau_\eta)^{1/2}}\frac{1}{(v+s_1+|\tau|+\tau_\eta)^{3/2}}ds_1dv\\
&+\frac{1}{2T}e^{-\frac{|\tau|}{T}}\int_{s_1=0}^{\infty}\int_{v=0}^{\infty}\int_{h=s_1+|\tau|}^{s_1}e^{-\frac{2s_1-h}{T}}\frac{1}{(v+\tau_\eta)^{1/2}}\frac{1}{(v+h+\tau_\eta)^{3/2}}ds_1dhdv.
\end{align*}
The first term on the RHS of this equality cancels an identical term of opposite sign in \refp{eq:covarXtau1}. It is then not difficult to show that the two others terms on the RHS remain bounded, for a given $|\tau|>0$, when $\tau_\eta\to 0$. We have thus demonstrated that the covariance of $X_{\tau_\eta}$ remains bounded at infinite Reynolds number, and we designate
\begin{align*}
\E \left[X(t)X(t+\tau)\right]&=\lim_{\tau_\eta\to 0}\E \left[X_{\tau_\eta}(t)X_{\tau_\eta}(t+\tau)\right]\\
&= \frac{1}{2}\int_{s_1=0}^{\infty}\int_{v=0}^{\infty}e^{-\frac{s_1}{T}}\frac{1}{\sqrt{v}}\frac{1}{(v+s_1+|\tau|)^{3/2}}ds_1dv\\
&+\frac{1}{2T}e^{-\frac{|\tau|}{T}}\int_{s_1=0}^{\infty}\int_{v=0}^{\infty}\int_{h=s_1+|\tau|}^{s_1}e^{-\frac{2s_1-h}{T}}\frac{1}{\sqrt{v}}\frac{1}{(v+h)^{3/2}}ds_1dhdv.
\end{align*}
The integral over $v$ may be carried on with the exact result provided in \refp{eq:ExactIntPL}, which brings us to the simplified expression
\begin{align*}
\E \left[X(t)X(t+\tau)\right]&= \int_{s=0}^{\infty}e^{-\frac{s}{T}}\frac{1}{s+|\tau|}ds+\frac{1}{T}e^{-\frac{|\tau|}{T}}\int_{s=0}^{\infty}\int_{h=s+|\tau|}^{s}e^{-\frac{2s-h}{T}}\frac{1}{h}dsdh.
\end{align*}
Performing an integration by parts over $h$, it is easy to justify that the second term in the RHS of the former equality remains bounded when $\tau\to 0$. As for the first term, it diverges when $\tau\to 0$ and will thus dominate the covariance of $X$ at small scales. For these reasons, we finally deduce the following equivalent of the covariance function at small scales:
\begin{align*}
\E \left[X(t)X(t+\tau)\right]\build{\sim}_{\tau\to 0}^{}\int_0^{\infty}e^{-\frac{s}{T}}\frac{1}{s+|\tau|}ds.
\end{align*}
A similar integral has been encountered in section \ref{Sec:AsyVarRe}, where it is shown that it diverges with $\tau$ according to $\log(T/|\tau|)$, which entails the proposition \refp{eq:ResCovarAsymptTau}.

\subsection{A causal multifractal process for pseudo-dissipation and its statistical properties} \label{Ann:CausMulAnalytic}

In the spirit of \cite{Sch03}, we consider the process
\begin{align}\label{eq:ProcYtau}
Y_{\tau_\eta}(t) = \sqrt{\mu^l} X_{\tau_\eta}(t) - \frac{\mu^l}{2}\E \left[\left(X_{\tau_\eta}\right)^2\right]
\end{align}
which contains the variance of $X_{\tau_\eta}$ in the stationary regime \refp{eq:VarXtau} and whose dynamics is given by
\begin{equation}\label{eq:sdeY}
dY_{\tau_\eta} = \left[-\frac{1}{T} \left(Y_{\tau_\eta}(t)+\frac{\mu^l}{2}\E \left[\left(X_{\tau_\eta}\right)^2\right]\right)+\sqrt{\mu^l}\beta_{\tau_\eta}(t)\right]dt + \sqrt{\frac{\mu^l}{\tau_\eta}}W(dt).
\end{equation}
The respective Lagrangian multiplicative chaos, which is causal and stationary, is readily obtained while exponentiating the Gaussian process $Y_{\tau_\eta}$:
\begin{equation}\label{eq:Phi}
\varphi (t)= \frac{1}{{\tau_\eta}^2}e^{Y_{\tau_\eta}(t)}.
\end{equation}
An application of Ito's lemma leads to the stochastic dynamics of the pseudo-dissipation as seen by a Lagrangian particle along its trajectory, namely
\begin{align}\label{eq:sdePhi}
d\varphi &= \varphi(t)\left[-\frac{1}{T} \left(\log [\tau_\eta^2\varphi(t)]+\frac{\mu^l}{2}\E \left[\left(X_{\tau_\eta}\right)^2\right]\right)+\sqrt{\mu^l}\beta_{\tau_\eta}(t) +\frac{\mu^l}{2{\tau_\eta}}\right]dt+ \sqrt{\frac{\mu^l}{\tau_\eta}}\varphi(t)W(dt),
\end{align}
that will eventually reach a stationary regime for any bounded, non vanishing and positive initial condition.

From \ref{eq:Phi} we extract the statistical properties of $\varphi$. First, for positive integers $q$, its moments of order $q$ are given by
$$ \E \left[\varphi^q\right] = \frac{1}{{\tau_\eta}^{2q}}e^{\frac{\mu^l}{2}q(q-1)\E \left[\left(X_{\tau_\eta}\right)^2\right]}.$$
As discussed in section \ref{Sec:AsyVarRe}, the variance of $X_{\tau_\eta}$ diverges logarithmically with $\tau_\eta$. Assuming that the sub-leading term is constant, call it $\tilde g(0)$, we may write, when $\tau_\eta\to 0$,
$$\E \left[\left(X_{\tau_\eta}\right)^2\right] = \log\left(\frac{T}{\tau_\eta}\right)+\tilde g(0)+o(1),$$
such that
\begin{equation}\label{eq:AnnMomqPhitaueta}
\E \left[\varphi^q\right] \build{\sim}_{\tau_\eta\to 0}^{}\frac{1}{{\tau_\eta}^{2q}}e^{\frac{\mu^l}{2}q(q-1)\tilde g(0)}\left(\frac{T}{\tau_\eta}\right)^{\frac{\mu^l}{2}q(q-1)}.
\end{equation}
In a similar fashion, let us now compute the moments of the averaged pseudo-dissipation $\varphi_\tau (t)$ over a time interval $\tau$ \refp{eq:MeanTauDiss}. We get
$$ \E \left[\varphi_\tau^q\right]=\frac{1}{{\tau_\eta}^{2q}}\frac{1}{\tau^q}\int_{[t-\tau,t]^q}e^{\mu^l\sum_{i<j}^q\E\left[X_{\tau_\eta}(s_i)X_{\tau_\eta}(s_j)\right]}\prod_{i=1}^qds_i.$$
We have seen in section \ref{Ann:AsymptCovXtaueta} that the covariance of $X_{\tau_\eta}$ remains bounded for non vanishing time lag when $\tau_\eta\to 0$. Furthermore, in this limit of infinite Reynolds number, we have also seen that this asymptotical covariance diverges logarithmically at the origin. Assume now that the sub-leading terms remain bounded and call $\tilde g(t)$ such a bounded function. We can then write
$$\lim_{\tau_\eta\to 0}\E\left[X_{\tau_\eta}(0)X_{\tau_\eta}(t)\right] =\log^+\left(\frac{T}{|t|}\right)+\tilde g(t),$$
as it was found in the Eulerian context \refp{eq:LogCorrXEta}. It is then not difficult to show that $ \tau_\eta^{2q}\E \left[\varphi_\tau^q\right]$ remains bounded when $\tau_\eta\to 0$ for $q<1+2/\mu^l$ to obtain the following behaviour at small scales:
\begin{align}\label{eq:AsymptPhitauq}
\lim_{\tau_\eta\to 0}\tau_\eta^{2q}\E \left[\varphi_\tau^q\right]&=\frac{1}{\tau^q}\int_{[0,\tau]^q}\prod_{i<j}^q\left( \frac{T}{\min (|s_i-s_j|,T)}\right)^{\mu^l}e^{\mu^l\sum_{i<j}^q F(s_i-s_j)}\prod_{i=1}^qds_i\notag\\
&\build{\sim}_{\tau\to 0}^{}\left( \frac{T}{\tau}\right)^{\frac{\mu^l}{2}q(q-1)}e^{\frac{\mu^l}{2}q(q-1) \tilde g(0) }\int_{[0,1]^q}\prod_{i<j}^q\frac{1}{|s_i-s_j|^{\mu^l}}\prod_{i=1}^qds_i.
\end{align}
Note that the condition on $q$, that is $q<1+2/\mu^l$, ensures that the integrals in \refp{eq:AsymptPhitauq} exist.

%%%%%%%%%%%%%%%%%%%%%%%%%%%%%%%%%%%%%%%%%%%%%%%%%%%%%%%%%%%%%%%%%%%%%%%%%%%%%%%%%%%%%%%%%%%%%%%%%%%%%%%%%%%%%%%%%%%%%%%%%%%%%%%%

\bibliographystyle{jfm}

\bibliography{./mybiblioJune2015}

\end{document}